\documentclass[journal]{IEEEtran}
\usepackage{graphicx}

\usepackage{amssymb}
 \usepackage[utf8]{inputenc}
 \usepackage{newunicodechar}
\newunicodechar{ﬁ}{fi}
\usepackage{tipa}
\usepackage{float}
\usepackage{subfig}
\usepackage{amsmath}
\usepackage[usenames, dvipsnames]{color}

\usepackage[capitalize,nameinlink]{cleveref}

\usepackage{mathtools}  
\usepackage{amsmath}
\usepackage{amssymb}
\usepackage{tabulary}
\usepackage{booktabs}

\ifCLASSINFOpdf
\else
\fi
\usepackage{amsmath,amssymb,amsthm}
\usepackage{enumerate}
\usepackage{color}
\usepackage[mathscr]{euscript}
\usepackage{url}
\usepackage[all]{xy}
\usepackage{booktabs}
\usepackage{graphicx}
\usepackage{epstopdf}
\usepackage{algorithmic}

\newcommand{\T}{{\sf T}}        % transposition

\newcommand{\vect}[1]{\boldsymbol{#1}}

\definecolor{darkred}{rgb}{0.7,0,0}
\definecolor{darkgreen}{rgb}{0,0.46,0}
\definecolor{purple}{rgb}{0.6,0,0.5}

\begin{document}
% \title{A New Re-synchronization Method based Multi-modal Fusion for Automatic Continuous Cued Speech Recognition}
\title{Re-synchronization using the Hand Preceding Model for Multi-modal
Fusion in Automatic Continuous Cued Speech Recognition}

\author{Li~Liu,~\IEEEmembership{Member,~IEEE,}
        Gang~Feng,
        Denis~Beautemps,
        and~Xiao-Ping~Zhang,~\IEEEmembership{Senior Member,~IEEE}
\thanks{L. Liu is with Shenzhen Research Institute of Big Data, Shenzhen, China. E-mail: liliu.math@gmail.com. G. Feng and D. Beautemps are with Univ. Grenoble Alpes, CNRS, Grenoble INP, GIPSA-lab, 38000 Grenoble, France. E-mail: gang.feng@gipsa-lab.grenoble-inp.fr, Denis.Beautemps@gipsa-lab.grenoble-inp.fr. X.-P. Zhang is with the Dept. of Electrical, Computer and Biomedical Engineering, Ryerson University, Toronto, ON, Canada. E-mail: xzhang@ee.ryerson.ca.}
\thanks{L. Liu and X.-P. Zhang are the corresponding authors.}
\thanks{Part of this work has been presented in conference Eusipco 2019.}}

\maketitle

\begin{abstract}
Cued Speech (CS) is an augmented lip reading system complemented by hand coding, and it is very helpful to the deaf people.
Automatic CS recognition can help communications between the deaf people and others.
Due to the asynchronous nature of lips and hand movements,
fusion of them in automatic CS recognition is a challenging problem. In this work, we propose a novel re-synchronization procedure for multi-modal fusion, which aligns the hand features with lips feature. It is realized by delaying hand position and hand shape with their optimal hand preceding time which is derived by investigating the temporal organizations of hand position and hand shape movements in CS. 
This re-synchronization procedure is incorporated into a practical continuous CS recognition system that combines convolutional neural network (CNN) with multi-stream hidden markov model (MSHMM). A significant improvement of about 4.6\% has been achieved retaining 76.6\% CS phoneme recognition correctness compared with the state-of-the-art architecture (72.04\%), which did not take into account the asynchrony issue of multi-modal fusion in CS. 
To our knowledge, this is the first work to tackle the asynchronous multi-modal fusion in the automatic continuous CS recognition. 
\end{abstract}


\IEEEpeerreviewmaketitle

\section{Introduction}
\label{sec:Introduction}
\IEEEPARstart{C}{ommunication} is one of the most important parts of human life, and more and more attention has been paid to improve communications among the disabled people in nowadays' society. It was reported by the \textit{World Health Organization} (WHO) \cite{whodeaf} that more than 5\% of the world’s population (466 million people) has disabling hearing loss (432 million adults and 34 million children) in the world. As one of the most common communication ways for deaf people, lip reading \cite{dodd1987hearing,nicholls1982cued} helps the deaf or hearing impaired people access spoken speech, and has undoubtedly improved communication of these people a lot. However, a significant drawback is that lip reading can only provide insufficient information in most cases. In fact, it cannot distinguish some contrasts, for example [p] vs. [b], which is caused by the similarity of labial shapes. As a result, this problem makes it difficult for deaf or hearing impaired people to access speech only by lip reading. 

\begin{figure*}[htb]
\begin{minipage}[b]{1.0\linewidth}
 \centering
 \centerline{\includegraphics[width=13.5cm]{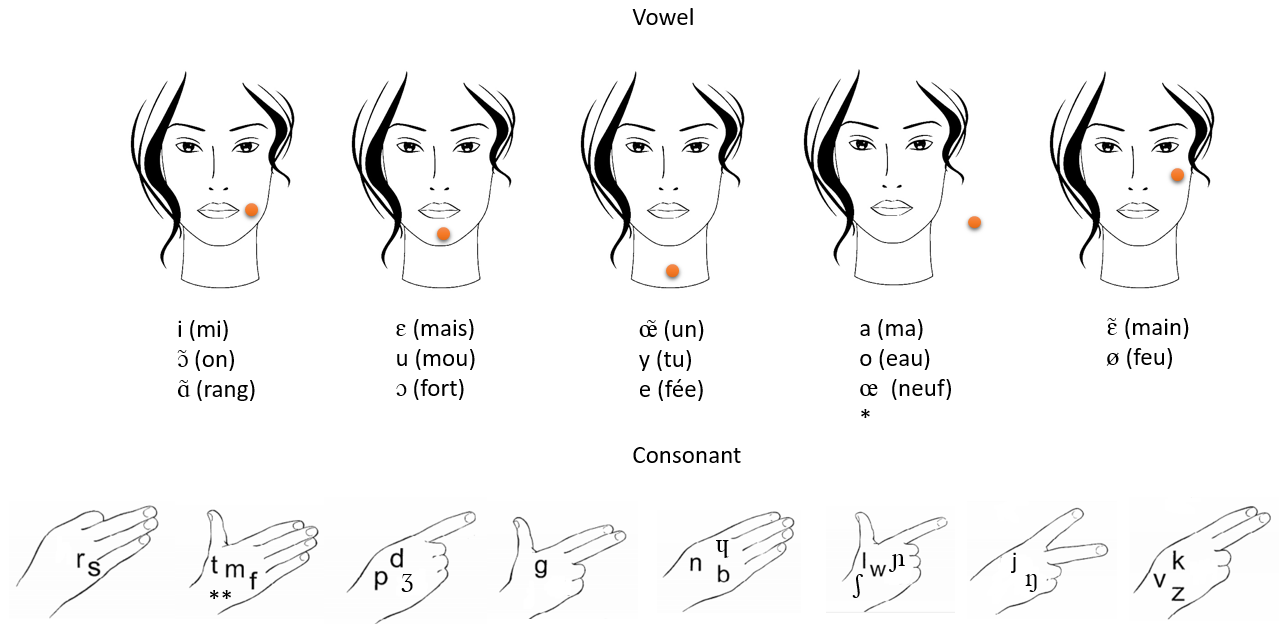}}
\end{minipage}
\caption{Hand coding in French Cued Speech system. Five hand positions (\emph{mouth, chin, throat, side, cheek}) are used to code vowel groups. Eight hand shapes are used to code consonant groups. The * in the \emph{side} position is used to code a single consonant, and the ** in the full hand shape is used to code a single vowel.}
\label{fig:CS_hengxiang}
\end{figure*}

To overcome insufficient information of lip reading and improve reading ability of deaf children, in 1967, Cornett \cite{cornett1967cued} invented the first \emph{Cued Speech} (CS) system for the American English, which complements lip reading and makes all the phonemes of a spoken language
clearly visible. This system is based on four hand positions and eight hand shapes in which two major criteria were set: the minimum effort for encoding, and the maximum visual contrast. In the French CS named \textit{Langue française Parlée Complétée (LPC)} \cite{lasasso2010cued} (see \cref{fig:CS_hengxiang}), five hand positions (i.e., \emph{mouth, chin, throat, side, cheek}) are used to encode vowel groups, and eight hand shapes are used to encode consonant groups. For British English CS, four hand positions are used to code monophthong vowel groups and eight hand shapes are designed to code consonant groups. By using CS systems, sounds that may look identical on lips (e.g., /y/, /u/ and /o/), can be distinguished using hand information, and thus it is possible for the deaf people to understand a spoken language using visual information alone.

CS has drawn increasing attention from all over the world and has been adapted to more than sixty spoken languages for the moment. Another widely used communication method in deaf community is \emph{Sign language} (SL) \cite{stokoe2005sign, liddell1989american, valli2000linguistics} which was developed in the early 18th century. SL is a language with its own grammar and syntax, while CS is a visual representation of spoken languages. The deaf people are able to master and learn the native language more easily by using CS \cite{lasasso2010cued,reynolds2007examination}. 

This work investigates the framework of asynchronous multi-modal fusion \cite{tang2017integration}, \cite{sun2019rtfnet,sun2017improving,sun2018motion,sun2019active} applied to automatic continuous CS recognition. Since the multi-modal streams in CS (i.e., lips, hand shape and hand position) that need to be fused are naturally asynchronous \cite{attina2004pilot, attina2005temporal}, the feature fusion of CS recognition posses a major challenge. The supervised CS recognition system will be interfered if the lips and hand information are misaligned. 

In the existing literature, CS feature modalities are assumed to be synchronous in the recognition task. In \cite{heracleous2010cued,heracleous2012continuous}, direct feature fusion (i.e., direct concatenation of the features) was applied to the isolated\footnote{Isolated CS recognition means that the temporal segmentation of each phoneme to be recognized are given at the test stage.} CS recognition without taking into account the asynchrony issue. 
In the state-of-the-art \cite{liu2018interspeech}, 
a tandem architecture that combines \emph{convolutional neural network} (CNN) \cite{lecun2015deep, goodfellow2016deep} with \emph{multi-stream hidden markov model} (MSHMM) \cite{potamianos2003recent} was used for the continuous CS recognition, and is referred as $\mathcal{S}_3$ in this work. In \cite{liu2018interspeech}, MSHMM merges different features by giving weights to three feature modalities of CS, but it does not take into account the asynchrony issue between them. Therefore, there is still room for improvement regarding the CS recognition performance by exploring a pre-processing approach to tackle the fusion of asynchronous multi-modalities. 

\begin{figure*}[htb]
\begin{minipage}[b]{1.0\linewidth}
\centering
\centerline{\includegraphics[width=11.5cm]{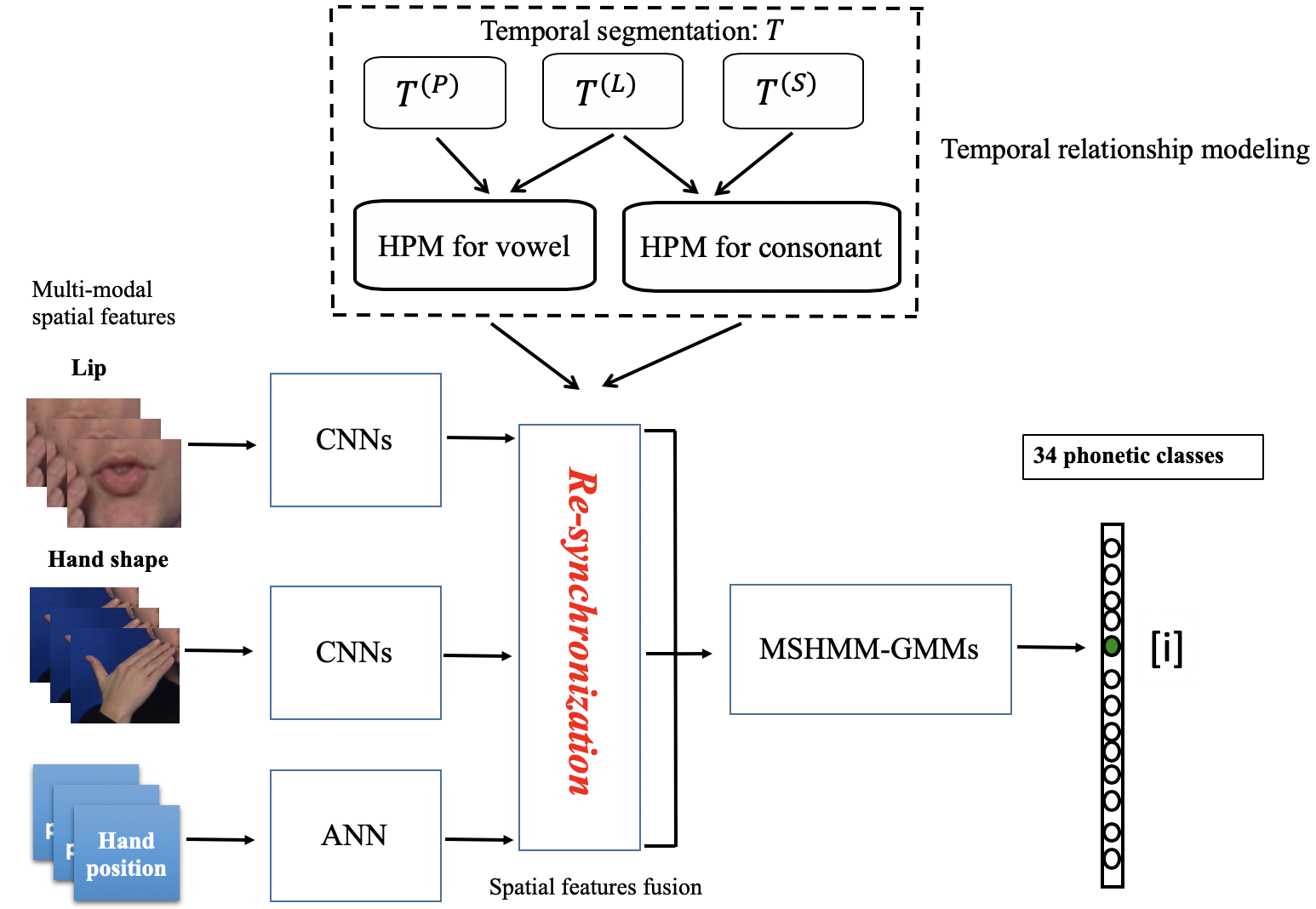}}
\end{minipage}
\caption{Overview of the proposed automatic CS recognition architecture $\mathcal{S}_{\rm re}$ in this work with the  re-synchronization procedure. 
${\rm T}^{\rm (L)}, {\rm T}^{\rm (P)},{\rm T}^{\rm (S)}$ are the temporal segmentations/boundaries for phonemes in case of lips, hand position and hand shape, respectively. Compared with the state-of-the-art architecture $\mathcal{S}_3$ in \cite{liu2018interspeech}, the temporal relationship modeling (dotted box) and the re-synchronization procedure (i.e., the tier before the MSHMM-GMM decoder) are added.}
\label{fig:HMM_architecture_resychrony}
\end{figure*}

In this work, we propose a new architecture based on a novel re-synchronization method for asynchronous multi-modal feature fusion in CS. The novelty stems in investigating the temporal organization of hand movement, which allows us to obtain the optimal \emph{hand preceding time} (HPT) for both vowels and consonants. More precisely, our method composes of two main stages.
\begin{enumerate}

    \item  First of all, we build a \emph{hand preceding model} (HPM) by analyzing the HPT on the database containing four CS speakers, and show that this model provides efficient segmentations of hand movements for all four speakers. Our proposed model significantly improves hand position recognition accuracy, and present the first main contribution of this work.
    \item Secondly, we propose an efficient re-synchronization procedure based on the HPM to align the hand feature stream, so that the hand and lips movements are statistically synchronous, providing a good fusion condition. By incorporating this re-synchronization procedure into the tandem CNN-MSHMM architecture (see $\mathcal{S}_{\rm re}$ in \cref{fig:HMM_architecture_resychrony}), the automatic continuous CS recognition obtains a significant improvement compared with the $\mathcal{S}_{3}$ in the state-of-the-art \cite{liu2018interspeech}. This is the second main contribution of the present work. To our knowledge, the proposed re-synchronization procedure is the first method to tackle the asynchrony issue of the CS multi-modal fusion applied to the automatic continuous CS recognition.
\end{enumerate}

\section{Related works}
\label{subsec: Related works}
The literature on automatic CS recognition can be classified into three main categories: feature extraction, multi-modal temporal modeling and CS recognition modeling. We will discuss them separately in this section.

\subsection{Feature extraction in CS}
In the literature on the automatic CS recognition, video images were recorded with artifices applied to the CS speaker (blue sticks on the lips, blue marks on the hand and forehead) to mark the pertinent information and make their further feature extraction easier.
For example, in \cite{heracleous2010cued, heracleous2012continuous}, lips feature was extracted by tracking the color marks on speaker's lips. Then a threshold was applied to the gray level images to segment the blue lips. 
The coordinates of the color marks on the finger were used as features for the hand shape and hand position modeling.
In \cite{burger2005cued}, the speaker wore a one-colored glove in order to help the hand segmentation. 
Stillitano et al. \cite{stillittano2013lip} used active contours combined with parametric models to extract the lips contour in CS.

In our recent works \cite{liu2017tracking,liu2017inner,liu2017innerclnf}, several methods to get rid of these artifices on the speaker's lips and hand were explored.
A modified \emph{constrained local neural fields} (CLNF) model was proposed to extract the inner lips height and width, and an adaptive ellipse model was proposed for inner lips parameters extraction in CS. 
In this work, we adopt the deep CNN for the feature extraction of lips and hand shape, and use the \emph{artificial neural network} (ANN) \cite{lecun2015deep, svozil1997introduction} to process the hand position feature.

\subsection{Multi-modal temporal modeling in CS}
The temporal organization of hand movements in CS was studied in \cite{attina2004pilot,aboutabit2006hand,liu2019novel}.
For CV syllables\footnote{syllables consist of a consonant followed by a vowel.},
it was found that the hand position reaches its
target before the vowel being visible at lips  on average $239$ms \cite{attina2004pilot} (based on the non-sense syllables logatome, like `mamuma'), or $144.19$ms \cite{aboutabit2006hand} (based on the syllables extracted from French continuous sentences).
In this work, we focus on not only the HPT of vowels, but also that of consonants.

In our previous work \cite{liu2018handpreceding}, the relationship between the HPT of vowels and their target time instant (i.e., position in time) was analyzed. 
It was found that HPT follows a Gaussian distribution that remains almost the same for all the instants of vowels, except a small time interval (about $1s$) before the end of each sentence. In this work, 
instead of following the piece-wise linear relationship, which gives different HPT for each vowel, we make a simple but efficient assumption that the mean value of the Gaussian distribution is suitable for all vowels. Moreover, based on the HPT for vowels in \cite{liu2018handpreceding}, we explore the optimal HPT of consonants, and then propose a re-synchronization procedure to align the hand features with lips feature for CS multi-modal feature fusion.

\subsection{CS recognition modeling}
The early work on the CS recognition is for the isolated vowel recognition in \cite{aboutabit2007reconnaissance, beautemps2007telma}.
Then, the isolated CS phoneme recognition was realized in \cite{heracleous2010cued}, 
and the continuous CS speech recognition based on an corpus of isolated words was conducted in \cite{heracleous2012continuous} using the context-independent HMM.
For the automatic continuous CS recognition based on a corpus of continuous sentences, a tandem CNN-HMM architecture extracting the CS feature from raw image was proposed in \cite{liu2018interspeech,liu2018modeling}. One-stream context-dependent HMM and MSHMM are both used for the multi-modal feature fusion, and MSHMM obtains better performance. 
However, none of them takes into account the asynchrony issue of the multi-modalities in CS. 

In this work, we propose a new automatic CS recognition architecture $\mathcal{S}_{\rm re}$ by adding a re-synchronization procedure to process the features extracted by CNNs and ANN before feeding them to the context-dependent MSHMM-GMM decoder. 
The experimental results show that this re-synchronization procedure significantly improves the CS recognition performance compared with both the state-of-the-art of isolated \cite{heracleous2010cued, heracleous2012continuous} and continuous CS recognition \cite{liu2018interspeech}.

\section{Temporal organization of hand movement}
\label{sec: Problem formulation_}

\begin{figure*}[htbp!]
\begin{minipage}[b]{1.0\linewidth}
 \centering
 \centerline{\includegraphics[width=11.5cm]{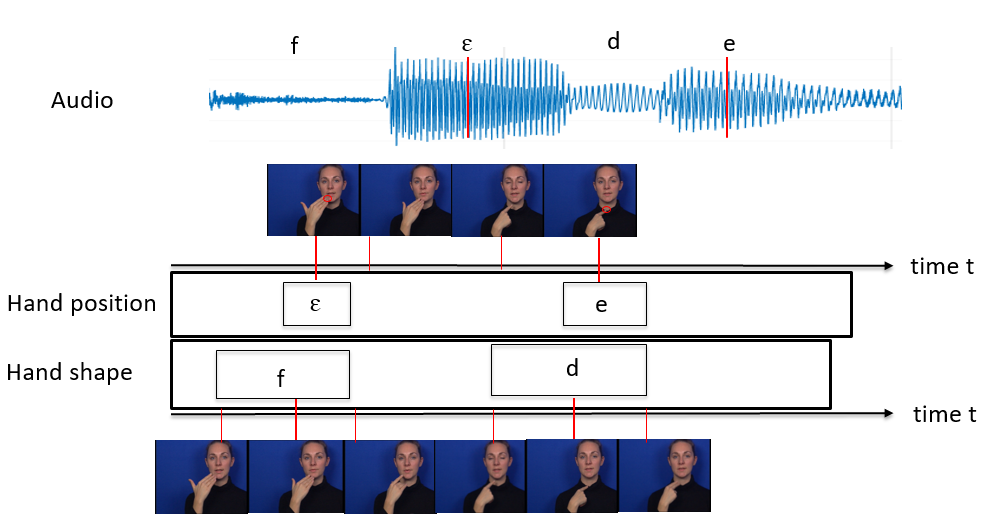}}
\end{minipage}
\caption{Illustration of the asynchrony phenomenon in the CS lips-hand movement. Two syllables [f\ \textipa{E}] and [d\ e] are extracted from the French sentence \textit{Il fait des achats}. The red vertical lines show the target instants of the vowels and consonants in the audio signal, hand position and shape streams, respectively.}
\label{fig:descychronization_hand_lips}
\end{figure*}

In this section, we investigate the temporal organization of hand movement in CS, which is important for establishing the HPMs for vowels and consonants in \cref{sec:A novel framework of the re-synchronization procedure}.
We first illustrate the \textit{hand preceding phenomenon} using an example of CS where the speaker utters \textit{fait des} ([f \textipa{E} d e]) in a French sentence \textit{Il fait des achats}.
It can be seen in \cref{fig:descychronization_hand_lips} that, when the CS speaker's hand points to the \emph{chin} position for the vowel [\textipa{E}], she is only lip-reading the consonant [f], and the vowel [\textipa{E}] has not begun yet.  
This phenomenon are also observed for the syllables [d e] in this example.

Let the rectangles on the 3rd row (hand position tier) denote the time intervals in which the hand reaches its target position, and the rectangles on the 4th row (hand shape tier) denote the time intervals in which the hand prepares its shape to indicate consonants.
It can be seen in  \cref{fig:descychronization_hand_lips} that the corresponding rectangles on the 3rd and 4th rows are aligned on the right end, and the bottom one is longer. 
The underlying reason is that in our database (will be introduced in \cref{subsubsec:Database}), we observe the following two facts.
(1) In the hand movement, the hand shape reaches its final shape at the same time as the hand position reaches its target for a vowel.
(2) The hand stays at the target position only for a small time interval. During this period, the hand shape is almost formed, but the hand continues moving and rotating. Therefore, these intervals are relatively long (about $200$ms). In fact, when the hand begins to leave the target position, the fingers move quickly to prepare the next hand shape.

As a consequence, the hand position is more sensitive to the asynchrony issue than the hand shape. This may be due to the intrinsic fact that the hand often stays in its target position for a relatively short time, while the full hand shape keeps a longer time in the coding process of CS. 
 
\begin{figure}[htbp!]%[h!]
\begin{minipage}[b]{1.0\linewidth}
\centering
\vspace{-0.2cm}
\centerline{\includegraphics[width=7.5cm]{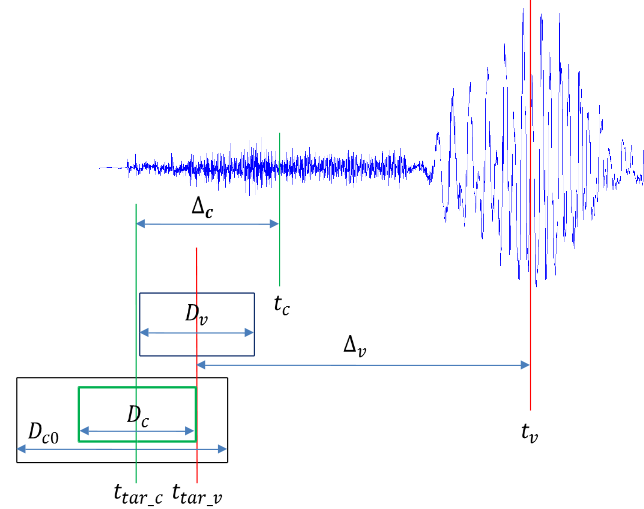}}
\end{minipage}
\caption{Illustration of different parameters concerning the hand movement temporal organization in CS. $\Delta_{\rm c}$ and $\Delta_{\rm v}$ are the HPT for consonants and vowels. $D_{\rm v}$ and $D_{\rm c}$ are the time intervals for target hand position and hand shape.}
\label{fig:figure_hand_mov_parameter}
\end{figure}

Now, we give some definitions and notations for HPMs. 
Without loss of generality, we only consider the CV syllables \cite{cornett1967cued}. 
We are interested in how long time the hand precedes the lips movement.
As shown in \cref{fig:figure_hand_mov_parameter},
for a vowel in a syllable, the middle instant of this vowel in the audio signal is denoted by $t_{\rm v}$, and the target instant in the hand position movement is denoted by $t_{\rm tar\char`\_v}$. $D_{\rm v}$ is the time interval in which the hand reaches its target position, which is set to be $60$ms experimentally. 
Then the HPT (in ms) for vowels is
\begin{equation}
\Delta_{\rm v} = t_{\rm v} - t_{\rm tar\char`\_v}.
\label{equ:hand_preceding_time}
\end{equation}

For a consonant in a syllable, the middle instant of this consonant in the audio signal is denoted by $t_{\rm c}$. Since the hand preceding phenomenon, $t_{\rm c}$ precedes $t_{\rm v}$. 
We assume that the complete hand shape is realized at the same time as the target position $t_{\rm tar\char`\_v}$. However, this complete hand shape does not correspond to a single instant but a certain time interval naturally. This time interval is before $t_{\rm tar\char`\_v}$ because after this moment, the hand shape begins to change immediately. Let $D_{\rm c}$ be the time interval corresponding to a given hand shape, and $D_{c_0}$ be the time interval in which a hand shape is almost formed but continues moving and rotating. The middle instant of $D_{\rm c}$ is denoted by $t_{\rm tar\char`\_c}$, which is the target instant in the hand shape. Then the HPT (in ms) for consonants is
\begin{equation}
\Delta_{\rm c}=t_{\rm c}-t_{\rm tar\char`\_c}.
\label{equ:hand_preceding_time_consonant}
\end{equation}

\section{Novel re-synchronization procedure}
\label{sec:A novel framework of the re-synchronization procedure}
In this section, we first formulate the problem of multi-modal fusion in CS recognition. Then, based on the temporal organization of hand movement in \cref{sec: Problem formulation_}, and by studying the HPT for vowels and consonants, we use them to explore the HPMs and build a new re-synchronization procedure.

\subsection{Problem formulation}
\label{subsec: Problem formulation}
In the automatic continuous CS phoneme recognition, features of lips $\vect{O}^{\rm (L)}$, hand position $\vect{O}^{\rm (P)}$ and hand shape $\vect{O}^{\rm (S)}$ are merged and fed into the phonetic decoder. Let the phoneme $\Upsilon$ extracted from a continuous French sentence at time $t$ be determined as
\begin{equation}
\Upsilon = \arg\max_{\Upsilon}P(\vect{O}^{\rm (LPS)}|\Theta_{\Upsilon}),
\end{equation}
where $\vect{O}^{\rm (LPS)} = [\vect{O}^{\rm (L)^\T}, \vect{O}^{\rm (P)^\T}, \vect{O}^{\rm (S)^\T}]^\T$ is the merged feature and $\Theta_{\Upsilon}$ is the model parameter for $\Upsilon$.

As we introduced in \cref{sec: Problem formulation_}, hand position and hand shape features are both asynchronous with lips feature in CS. Therefore, at time $t$, the direct concatenated feature will be interfered and not suitable to train one particular phoneme class $\Upsilon$. For example, in \cref{fig:descychronization_hand_lips}, the vowel [\textipa{E}] in lips feature is merged with [e] in hand position feature to train the reference vowel [\textipa{E}] if a direct concatenation fusion is applied to these features.

This work aims to propose a way to align $\vect{O}^{\rm (P)}$ and $\vect{O}^{\rm (S)}$ with $\vect{O}^{\rm (L)}$, i.e., to build transformations $\tau_{1}$ and $\tau_{2}$ such that
\begin{align}
        \vect{O}^{\rm (P)}_{\rm resy}=\tau_{1} (\vect{O}^{\rm (P)}),\\
        \vect{O}^{\rm (S)}_{\rm resy}=\tau_{2} (\vect{O}^{\rm (S)}),
        \label{equ:tao12}
\end{align}
are both synchronous with $\vect{O}^{\rm (L)}$. Then the merged feature $\vect{O}_{\rm resy}^{\rm (LPS)} = [\vect{O}^{\rm (L)^\T}, \vect{O}^{\rm (P)^\T}_{\rm resy},\vect{O}^{\rm (S)^\T}_{\rm resy}]^\T$ for phoneme $\Upsilon$ can be used to train the model of this phoneme without any interference. 

\subsection{Hand preceding model for vowels}
\label{subsec:Hand preceding model for vowel}
 
Now we establish the relationship between HPT and $t_{\rm v}$ based on our database (will be introduced in \cref{subsubsec:Database}). For this purpose, it is necessary to measure $t_{\rm tar\char`\_v}$. To make them precise enough, we determine them manually. A manual temporal segmentation of vowels and consonants for each sentence is accomplished by using the movie editor \textit{Magix} \cite{abreu2008podcasting, naylor2014magix}. 
Here, we set $t_{\rm tar\char`\_v}$ to be the middle instant of the temporal target interval, which contains several images around the hand target position.

\begin{figure*}
\centering
\subfloat[]{\includegraphics[width = 3.5in]{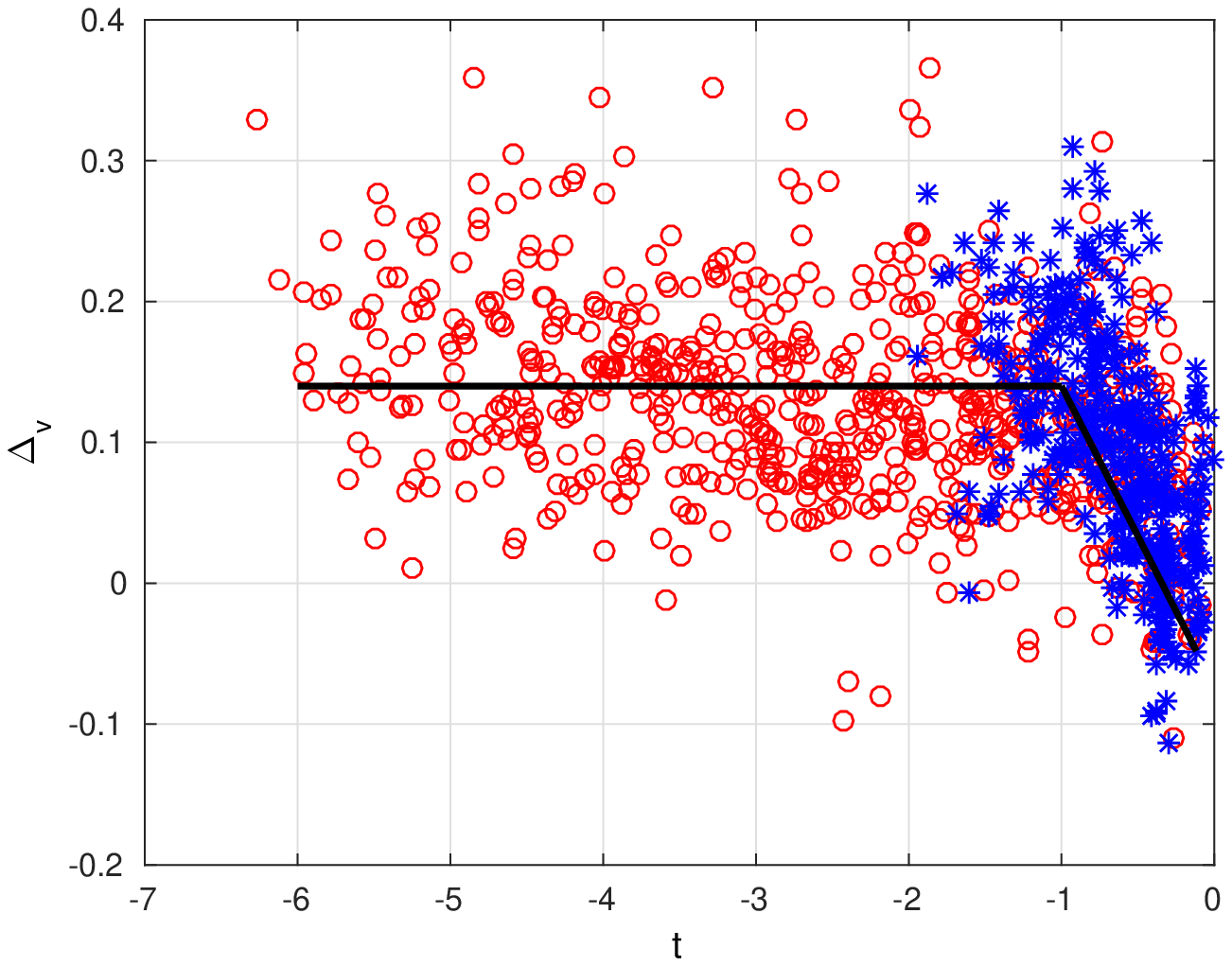}} 
\subfloat[]{\includegraphics[width = 3.5in]{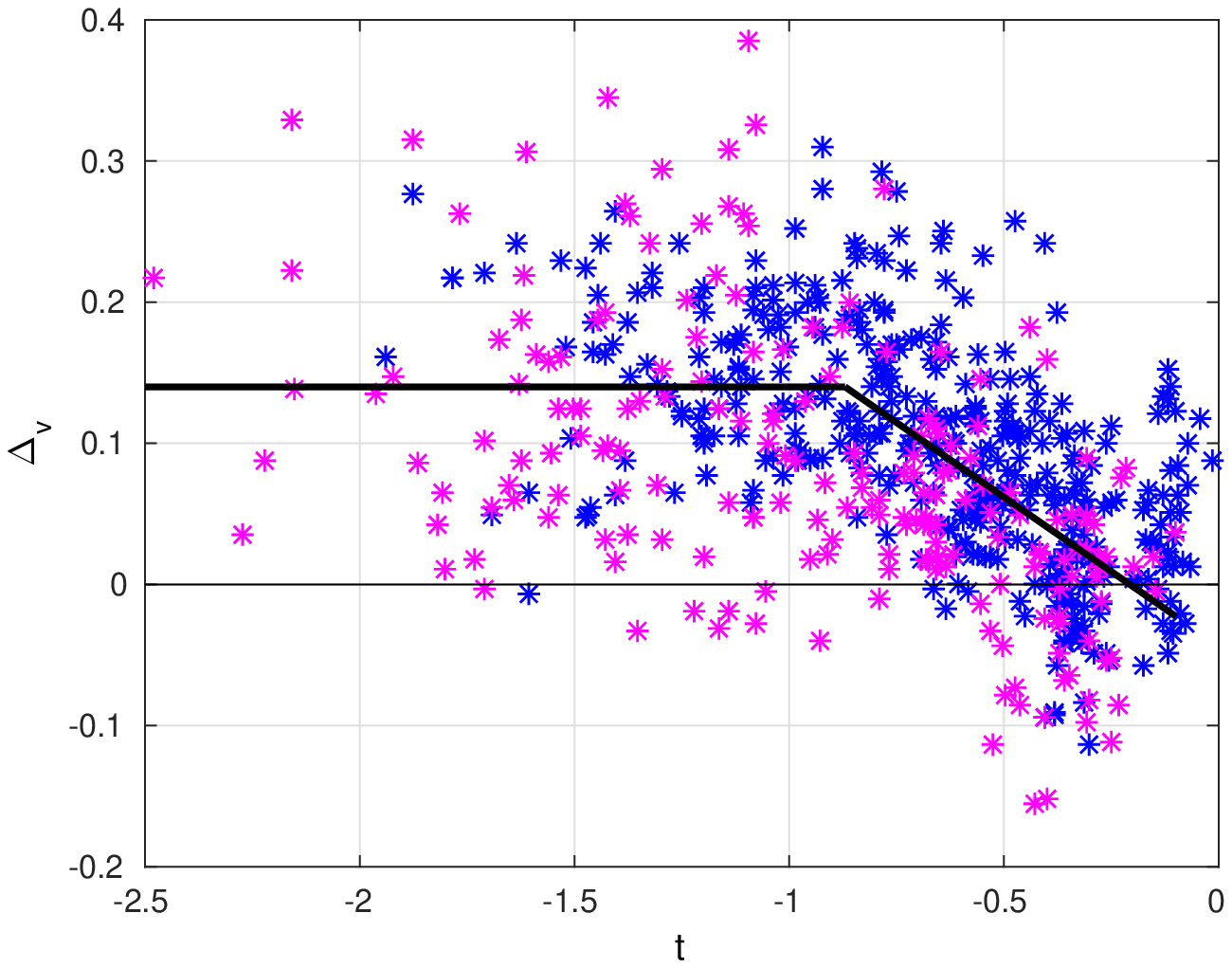}}\\
\subfloat[]{\includegraphics[width = 3.5in]{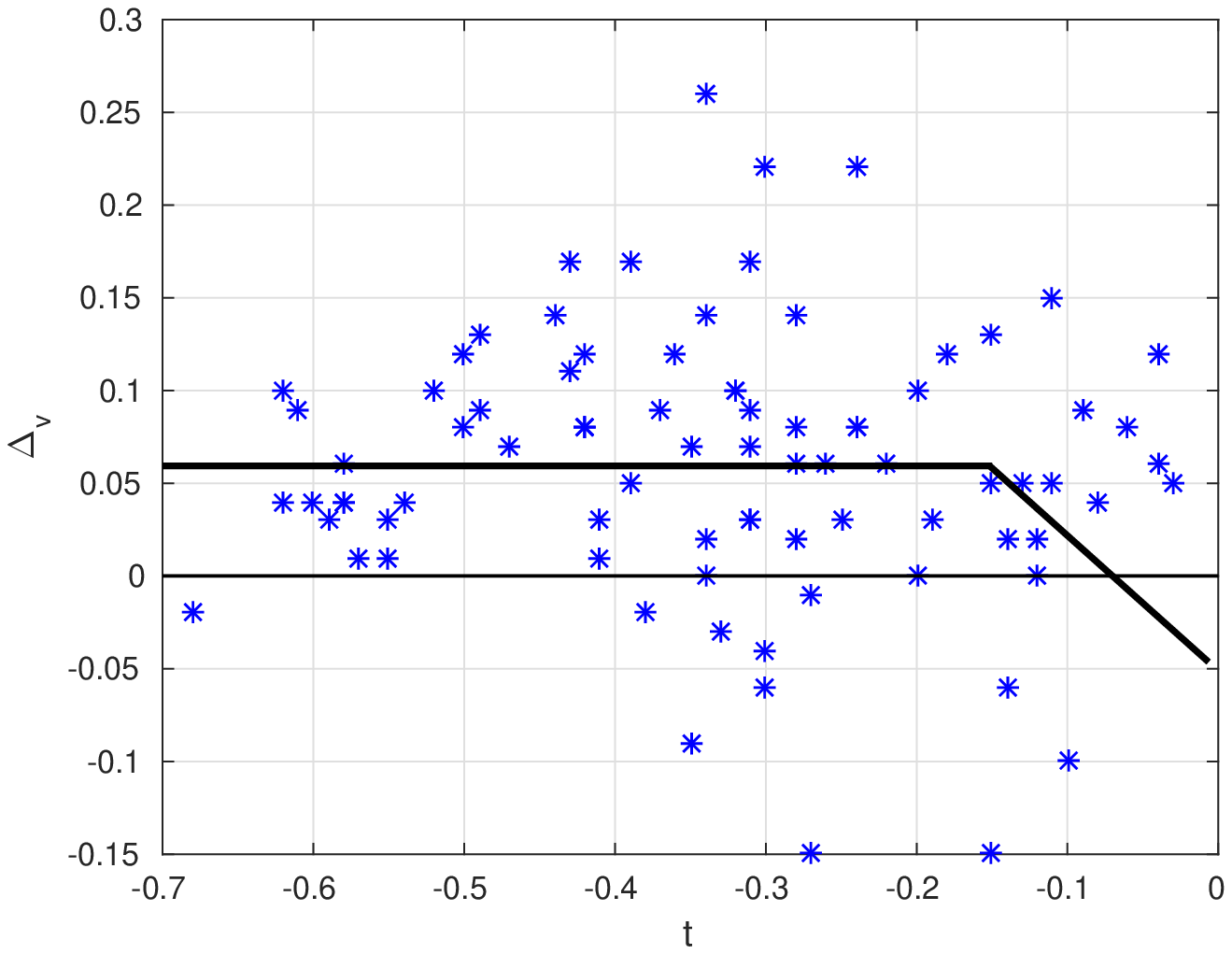}}
\subfloat[]{\includegraphics[width = 3.5in]{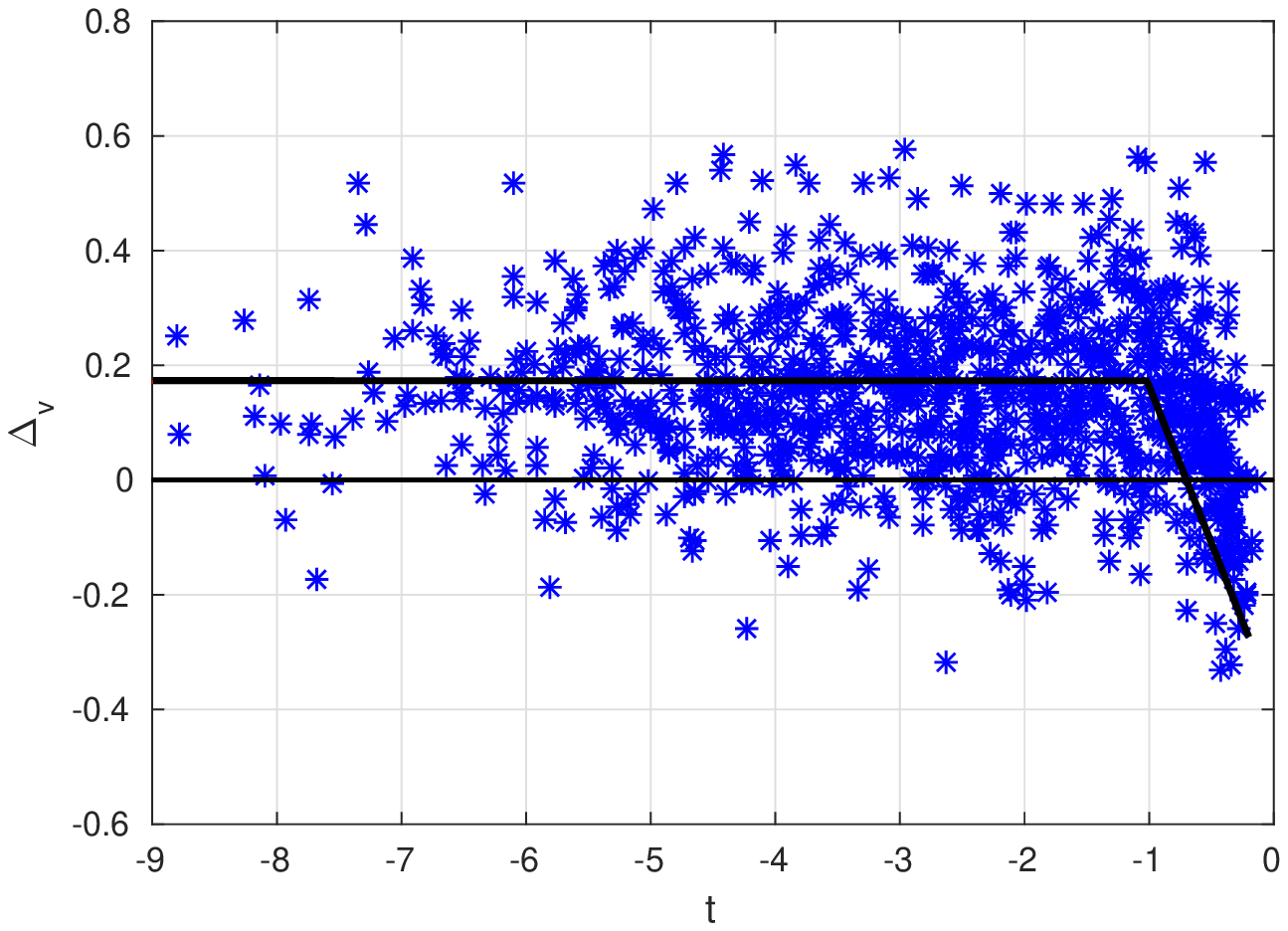}} 
\caption{HPT distribution and HPM. The abscissa is the vowel instant in sentences. All these points are aligned at the end of sentences, where the instant is 0. Y-axis: the $\Delta_{\rm v}$ (in seconds). In (a), the red circles show the distribution of
50 long sentences, and the blue stars show the distribution of 88 short sentences from \textit{LM} corpus. The black curve shows the HPM. In (b), the blue stars show the distribution of 88 short sentences from \textit{LM} corpus and the magenta stars show the distribution of 44 short sentences from \textit{SC} corpus. In (c), the blue stars show the distribution all 83 vowels of the \textit{MD} corpus made of 50 single words. In (d), the blue stars show the distribution all 1045 vowels of the \textit{CA} corpus made of 97 British English sentences.}
\label{fig:LM sabine}
\end{figure*}

All the $t_{\rm v}$ can be obtained by the audio-based segmentation. Then $\Delta_{\rm v}$ can be calculated by \eqref{equ:hand_preceding_time}. Taking the \textit{LM} speaker as an example, $\Delta_{\rm v}$ of 1066 vowels extracted from 138 sentences are plotted in \cref{fig:LM sabine}(a) with respect to $t_{\rm v}$. In this figure, all the points are aligned by their end (i.e., the instant 0) instead of the beginning, since in this way we can find a common rule between $\Delta_{\rm v}$ and $t_{\rm v}$ for all the vowels. 
More precisely, from the beginning of a sentence to a certain instant (about one second before the end), the statistical repartition of $\Delta_{\rm v}$ is almost the same in this period. Then, $\Delta_{\rm v}$ decreases until the end of the sentence and finally converges. By aligning all the sentences by their end, this phenomenon becomes very evident. Indeed, the distribution of short sentences and long sentences are superposed entirely at the end of sentences.
Besides, by comparing the $\Delta_{\rm v}$ distribution of other three speakers, we find that they follow the same repartition as shown in \cref{fig:LM sabine}(b)-(d).

Based on these observations, we build the HPM as
\begin{equation}
\Delta(t)=\left\{\begin{matrix}
 \overline{\Delta_{\rm v}} & 0<t<t_{0},\\ 
 at+b& t_{0}<t<L,
\end{matrix}\right.
\label{equ:linear}
\end{equation}
where $t$ is the time instant of vowel, and $\overline{\Delta_{\rm v}}$ is the mean value of the $\Delta_{\rm v}$ before the turning time instant $t_{0}$. As shown in \cref{fig:LM sabine}, we can see that the HPT of all vowels before $t_{0}$ follows a Gaussian distribution. Because we observe that the mean value $\overline{\Delta_{\rm v}}$ can reflect the main character of these points, we take it as the model. After $t_{0}$, we make a linear regression based on these points since the HPT for vowels decreases linearly. $a$ and $b$ in \eqref{equ:linear} are the slope and intercept of the linear line after $t_{0}$, respectively. In our case, $t_{0}$ stays at about $0.88s$ before the end of a sentence. The HPM \eqref{equ:linear} fits all sentences of four CS speakers.

The proposed HPM for the vowel will be evaluated in \cref{subsec:Evaluation of hand preceding model for hand position}, where a hand position recognition experiment is carried out.
In this experiment, we propose a temporal segmentation method for the hand position movement based on the HPM.
More precisely, from the audio-based segmentation, each temporal segmentation of vowels is shifted by $\Delta(t)$ using the HPM in \eqref{equ:linear} according to the vowel instant in the sentence.
Besides, the proposed temporal segmentation is also used to improve the efficiency of the ANN training for the hand position feature, as well as the CNN training for the hand shape feature (with respect to the HPM for the consonants) in the automatic CS recognition.

\subsection{Hand preceding model for consonants}
\label{subsec: Hand preceding model for consonants}
To investigate the HPM for the consonant, we first perform a statistical study on
\begin{equation}
\delta_{\rm cv}=t_{\rm v}-t_{\rm c},
\end{equation}
where $t_{\rm v}$ and $t_{\rm c}$ are the middle instant of the vowel and consonant in the audio signal.

Based on the \textit{LM} corpus, we randomly choose 10 sentences containing about 100 syllables.
$\delta_{\rm cv}$ of all these syllables are calculated and vary in a broad range. 
Therefore, we only consider its mean value (about 110ms).
Since the optimal $\Delta_{\rm v}$ is about $140$ms, from \cref{fig:figure_hand_mov_parameter}, we can deduce that $t_{\rm tar\char`\_v}$ precedes $t_{\rm c}$ by about 30ms (i.e., the difference between $140$ms and $110$ms). 
We assume that $D_{\rm c}$ is about $60$ms (i.e., three images) based on a large number of observations.
Consequently, $t_{\rm tar\char`\_c}$ precedes $t_{\rm c}$ by about $60$ms, which is an theoretical estimation of the optimal $\Delta_{\rm c}$ for all consonants. 

\begin{figure}[htbp!]
\begin{minipage}[b]{1.0\linewidth}
\centering
\centerline{\includegraphics[width=8.0cm]{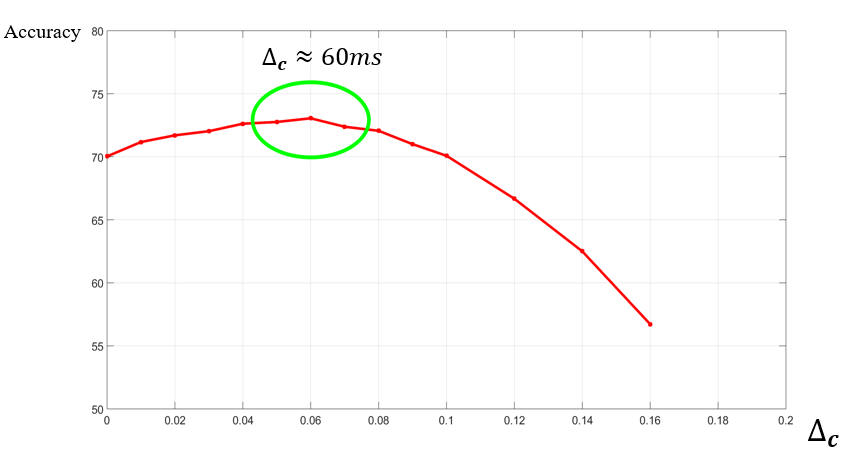}}
\end{minipage}
\caption{Hand shape recognition accuracy using different segmentations in function of $\Delta_{\rm c}$ (red curve). The green circle highlights the optimal recognition accuracy when $\Delta_{\rm c}\approx 60$ms.}
\label{fig:short_segmentation}
\end{figure}

Then we experimentally determine the optimal $\Delta_{\rm c}$. To do this, we perform a hand shape recognition experiment using the CNN-based hand features (will be introduced in \cref{subsec:Automatic continuous Cued Speech recognition}) and multi-Gaussian classifier. In the recognition, we apply several different temporal segmentations, which are derived by shifting the audio-based segmentation with different $\Delta_{\rm c}$ (from 0 to $160$ms with a step of $10$ms).
We denote by $\Delta_{\rm c}^{*}$ the $\Delta_{\rm c}$ that gives the best hand shape recognition accuracy.
A hand shape recognition experiment based on the \textit{LM} corpus of all 476 sentences is conducted. 80\% of the data is used for training and the rest 20\% is for test. The recognition results are shown in \cref{fig:short_segmentation}, which gives the recognition accuracy as a function of the monotonically increasing $\Delta_{\rm c}$. We observe a convex curve (red curve) with a local maximum value of about $60$ms. This is coherent with the theoretical analysis of $\Delta_{\rm c}$. Indeed, the peak region of this curve is relatively smooth, but the clear maximum value confirms that $\Delta_{\rm c}^{*}$ exists.

The temporal segmentation of hand shape is obtained using the same way as for the hand position. This temporal segmentation of hand shape will be used in the CNN training for hand shape feature extraction in \cref{subsec:Automatic continuous Cued Speech recognition}.

\subsection{Re-synchronization procedure}
\label{subsec:Two types of feature concatenations}

Instead of following the piecewise linear relationship in \eqref{equ:linear}, which gives different $\Delta_{\rm v}$ for the vowels after $t_{0}$,
we propose a simple but efficient assumption that $\Delta_{\rm v}^{*}$ 
is suitable for all vowels, as introduced in \cref{subsec:Hand preceding model for vowel}.
Similarly, we assume that $\Delta_{\rm c}^{*}$ is suitable for all consonants.

Now we propose the \emph{re-synchronization procedure} to align the hand shape and position features with lips feature based on the above HPMs for vowels and consonants. 
This procedure contains two steps:
\begin{enumerate}
\item Positively shift $\vect{O}^{\rm (P)}$ by $\Delta_{\rm v}^{*}$, i.e.,  
\begin{equation}
\vect{O}^{\rm (P)}_{\rm resy}(t) = \tau_{1} (\vect{O}^{\rm (P)})(t)= \vect{O}^{\rm (P)}(t-\Delta_{\rm v}^{*}),
  \label{equ:vowel_140}
\end{equation}
where $\Delta_{\rm v}^{*} = 140$ms.
\item Positively shift $\vect{O}^{\rm (S)}$ by $\Delta_{\rm c}^{*}$, i.e.,  
\begin{equation}
\vect{O}^{\rm (S)}_{\rm resy}(t) = \tau_{2} (\vect{O}^{\rm (S)})(t) = \vect{O}^{\rm (S)}(t-\Delta_{\rm c}^{*}),
 \label{equ:consonant_60}
\end{equation} 
where $\Delta_{\rm c}^{*} = 60$ms.
\end{enumerate}

\begin{figure*}[htbp!]%[h!]
\begin{minipage}[b]{1.0\linewidth}
\centering
\centerline{\includegraphics[width=11.5cm]{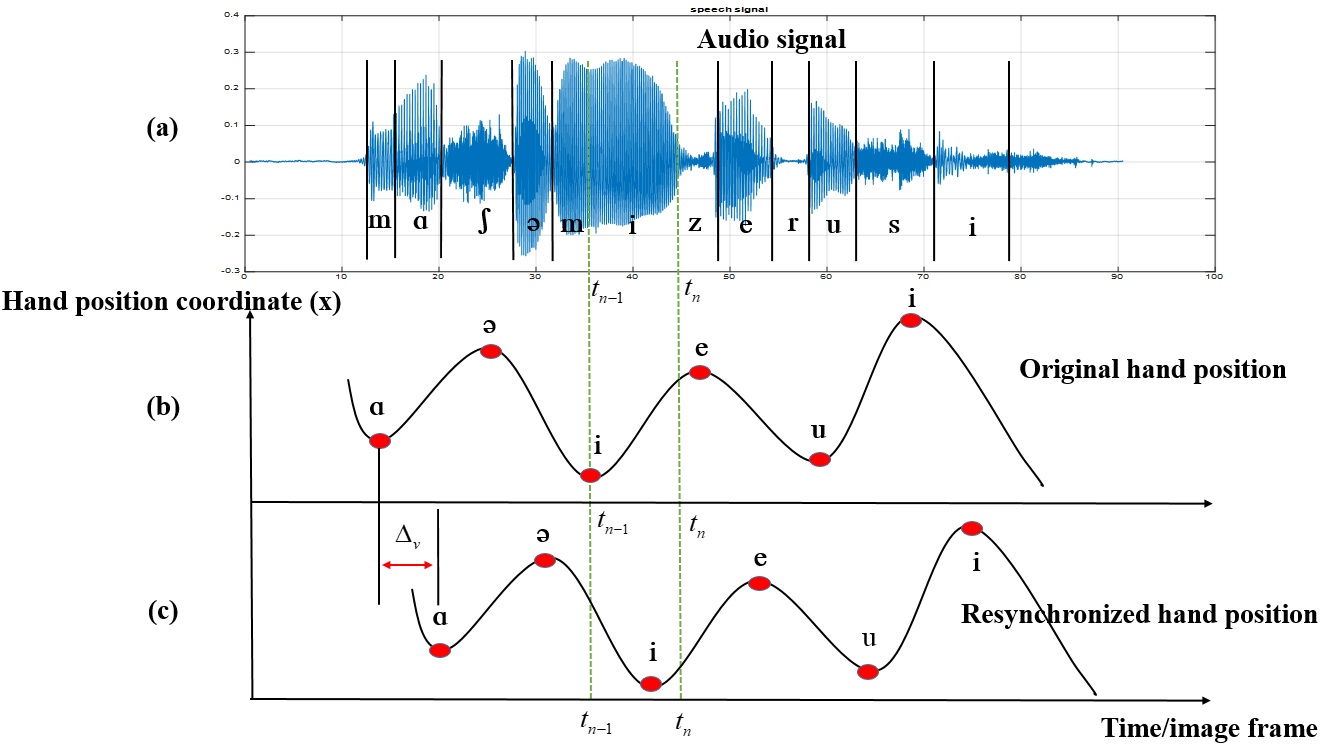}}
\end{minipage}
\caption{Illustration of the re-synchronization procedure. (a) The audio speech with its temporal segmentations and phonetic annotations. (b) The original hand position trajectory (i.e., $X$ coordinate of the hand back point). (c) The re-synchronized hand position derived by shifting the original hand position in (b) with $\Delta_{\rm v}$. Two green lines correspond to the audio based temporal segmentation of vowel [i].}
\label{fig:Aligned_hand_position_features}
\end{figure*}

We take the vowel case in \cref{fig:Aligned_hand_position_features} as an example to illustrate this procedure.
\cref{fig:Aligned_hand_position_features}(a) shows the audio signal of the French sentence \textit{Ma chemise est roussie} with the phonetic annotations. The lips feature is assumed to be synchronous with the audio signal \cite{tinwell2015effect}.
In \cref{fig:Aligned_hand_position_features}(b), the hand position is defined to be the $X$ coordinate of the hand back point. It is clear that the hand position stream is not synchronous with the audio signal. Therefore, a direct fusion of these two streams will cause some interference. In \cref{fig:Aligned_hand_position_features}(c), the aligned hand position stream is obtained by positively shifting the original one by $\Delta_{\rm v}^{*} = 140$ms as in \eqref{equ:vowel_140}. It turns out that the hand position stream is re-synchronized with the audio signal on average. 

For consonants, the alignment of the hand shape feature is similar, but with $\Delta_{\rm c}^{*} = 60$ms.
Even though the value of $\Delta_{\rm v}^{*}$ and $\Delta_{\rm c}^{*}$ may vary for different speakers, the proposed re-synchronization procedure is new, simple and efficient to improve the multi-modal fusion for continuous CS recognition. 

\section{Experimental setup}
\label{sec: Experiment setup}
In this section, we first introduce the database and the experimental metric for the HPMs as well as the CS continuous recognition experiments. Then technical details for the automatic continuous CS recognition are presented.

\subsection{Database}
\label{subsubsec:Database}
The database contains the recording data of three French CS speakers \textit{LM}, \textit{SC}, \textit{MD} and one British English CS speaker \textit{CA}.
Three French CS corpora were recorded in a sound-proof room of GIPSA-lab, France.
Color video images of the speaker's upper body are recorded at 50 fps, with a spatial resolution of 720$\times$576 pixels RGB images. 
The \textit{LM} speaker pronounces and codes a set of 238 French sentences in CS derived from a corpus described in \cite{aboutabit2007reconnaissance, gibert2005analysis}. Each sentence is repeated twice resulting in a set of 476 sentences (about $11770$ phonemes, and $120K$ images totally). The \textit{SC} corpus is made of 267 sentences from the same database as \textit{LM}. The \textit{MD} corpus contains videos of 50 French words made
of numbers and daily words. The corpus is uttered 10 times.
It should be mentioned that the \textit{LM} corpus is recorded without using any artificial mark, while the \textit{SC} and \textit{MD} corpora are earlier recorded with artificial marks \cite{aboutabit2006hand,aboutabit2007reconnaissance}. 
However, in this work, the \textit{SC} and \textit{MD} corpora are only used to establish and evaluate the HPM and they do not attend the automatic CS recognition.
Therefore, in this work, their artificial marks are not used at all although they exist.

The \textit{CA} corpus is the first British English CS corpus\footnote{This database will be made publicly available on Zenodo.},
which is recorded for this work in \emph{Cued Speech UK}\footnote{http://www.cuedspeech.co.uk/} association, without using any artificial mark. The professional CS speaker \textit{CA} (with no hearing impairment) is asked to simultaneously utter and encode a set of $97$ British English sentences (e.g., \textit{I feel it is a time to move to a new chapter in my career}). Color video images of the speaker's upper body are recorded at 25 fps, with a spatial resolution of $720$x$1280$. There are totally 907 monophthongs and 138 diphthongs in this corpus.

In \cref{sec:A novel framework of the re-synchronization procedure}, we take a subset of the whole database to build the HPM, which contains 138 sentences including 88 short sentences\footnote{In this work, a short sentence means a sentence with less than 4 vowels. Otherwise, it is a long sentence.} and 50 long sentences from \textit{LM} corpus (totally 1066 vowels), and 44 short sentences (196 vowels) from \textit{SC} corpus.
The hand position is manually determined in the rule that it is assumed to be the 2D position of the middle finger if the middle finger appears; otherwise, it is assumed to be the index finger.
Moreover, the hand back point is also tracked manually. The advantage of this point is that it is always visible in the CS coding process, and thus all the hand back points can be collected without any interruption. 

For the CS phoneme recognition, all 476 sentences of \textit{LM} corpus are used with 34 French phoneme classes ($14$ vowels and $20$ consonants). We have made this database publicly available\footnote{ https://doi.org/10.5281/zenodo.1206001.}. The French CS is described with $8$ lips visemes (as defined in \cite{aboutabit2007reconnaissance}), $8$ different hand shapes, and $5$ different hand positions (as defined in \cref{fig:CS_hengxiang}). The phonetic transcription is extracted automatically using Lliaphon software \cite{bechet2001lia} and post-checked manually to adapt it to the pronunciation of the CS speaker. 
The audio-based temporal segmentation of each vowel is extracted from the conventional \emph{audio speech recognition} (ASR) system in \textit{HTK 3.4} \cite{young1993htk}. Using forced alignment, the audio signal synchronous with the video is automatically labeled. 

\subsection{Experimental metric}
\label{subsec: Evaluation and metric}

For the experiments including the hand shape recognition (for investigating the $\Delta_{\rm c}^{*}$ in \cref{subsec: Hand preceding model for consonants}), the hand position recognition (for evaluating the HPM in \cref{subsec:Evaluation of hand preceding model for hand position}) and the CS vowel, consonant, phoneme recognition (for evaluating the proposed re-synchronization procedure in \cref{subsec: Evaluation of the re-synchronization procedure applied to the vowel and consonant recognition in Cued Speech} and \cref{subsec:Results of the automatic recognition using the re-synchronization procedure}), we randomly select 80\% of the sentences as the training set while the rest 20\% is the test set. Because each sentence is recorded twice, we allocate each sentence and its repetition either simultaneously to the training set or to the test set. All the results are repeated ten times with different training sets and test sets, so that the standard deviation (std) of these results can be controlled. It will be seen in \cref{sec:Experiment and Discussion} that the stds of all the experimental results are less than 0.5\%.

In CS phoneme recognition, the correctness of the MSHMM-GMM decoder
\begin{equation}\rm 
{\rm T}_{\rm corr} = (N-D-S)/N
\end{equation}
is chosen as the metric, where ${\rm N}$ is the number of phonemes in the test set, ${\rm D}$ is the number of deletion errors, and ${\rm S}$ is the number of substitution errors.

\subsection{Automatic continuous Cued Speech recognition}
\label{subsec:Automatic continuous Cued Speech recognition}
The continuous CS recognition (see \cref{fig:HMM_architecture_resychrony}) is carried out to evaluate the proposed re-synchronization procedure, and it contains three steps: feature extraction, multi-modal feature fusion and phonetic decoding. Note that the re-synchronization procedure is used to deal with the multi-modal feature fusion. Now we only introduce the methods for feature extraction and phonetic decoding.

\subsubsection{Cued Speech feature extraction}
\label{subsec:roi}
\paragraph{Lips and hand shape feature extraction}
We first extract the \emph{regions of interest} (ROIs) of lips and hand shape. The lips ROI is extracted using \emph{Kanade-Lucas-Tomasi} (KLT) feature tracker \cite{shi1994good} (with a dezooming process), and the hand shape ROI is extracted using \emph{Adaptive Background Mixture Model} (ABMM) \cite{stauffer1999adaptive}, which is able to track the variable-shape object. Then, lips and hand shape ROIs are transferred to 2D gray images and resized to a fixed size using the cubic interpolation.

\begin{figure*}[!t]
\centering
\makeatletter
\includegraphics[width=15.5cm]{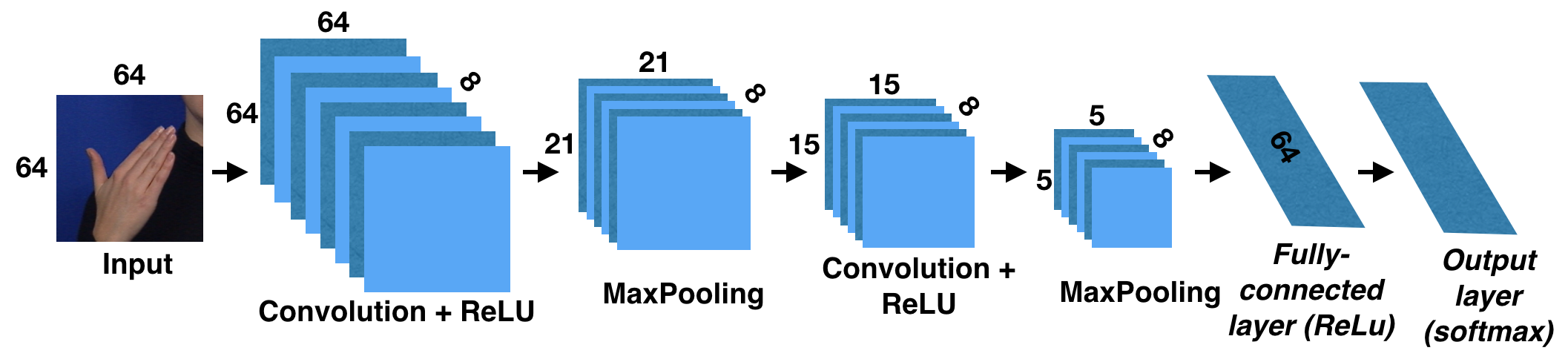}
\caption{Implementation details of the CNNs used to extract the hand shape features from hand ROI in CS.}
\label{fig:cnn_model_hand}
\end{figure*}

CNN is very powerful in extracting high-level features of images \cite{lecun2015deep}. In order to get rid of using the artifices (e.g., color marks on speaker's lips and hand), in this work, CNN is used to extract the lips and hand shape features from their raw ROI (i.e., nature images without marks). In this work, the CNN architecture in this work is almost the same as that in the state-of-the-art \cite{liu2018interspeech}. We investigated several architectures from scratch based on several convolutional/pooling layers, one or two fully connected layers, and one output
(i.e., softmax) layer, instead of using the well-known CNN backbones (e.g., ResNet, VGG). The cross-validation was used to optimize some hyper-parameters for each layer (i.e., the number of filters, the kernel size for the 2D convolutions, the down-sampling factor for the pooling layer, and the number of neurons in the fully connected layer). Finally, two convolutional layers with $8$ filters, a kernel size of $7$x$7$ pixels, a down-sampling factor of $3$ (in both vertical and horizontal directions), and $64$ hidden neurons in the fully connected layer were used. 

An ablation study was conducted concerning the number of CNN layers, adding batch normalization layer and changing the dropout probability. We investigated CNNs based on one to five convolutional/pooling layers. The recognition accuracies of the hand shapes and lips visemes by different CNNs are shown in \cref{tab:Ablation study of convolutional}. We can see that the best performance comes with two convolutional/pooling layers. Besides, the experimental results show no statistically significant difference by adding the batch normalization layer or changing the dropout probability (from 0.15 to 0.35).
\begin{table}[h!]
     \centering
     \small\addtolength{\tabcolsep}{-3pt}
      \caption{Ablation study on the number of CNN layers. Note that stds of the following results are less than 0.5\%.}  
        \label{tab:Ablation study of convolutional}
    \begin{tabular}{c c c c c c} 
     \hline\hline
        Accuracy(\%)/\#layers  &1  & 2&3  & 4&5  \\ [0.3ex] 
     \hline
     Lip visemes& 56.7 & 58.5 &58.1 &57.5 &57.0  \\
          \hline
    Hand shapes & 66.0 & 68.5 &67.6 &67.3 &66.5 \\
    \hline\hline
  \end{tabular}
\end{table}

At training stage, a mini-batch gradient descent algorithm based on the \textit{RMSprop} adaptive learning rate method (with a learning rate equal to $0.001$) was used to estimate the CNN parameters. The categorical cross-entropy was used as the loss function. Over-fitting was controlled using i) an early stopping strategy, i.e., 20\% of the training set was used as a validation set and the training was stopped when the error on this dataset stopped decreasing during $10$ epochs, and ii) a dropout mechanism (with a dropout probability of $0.25$). All models were implemented using the \textit{Keras} Python library \cite{chollet2015keras} and were trained using the Nvidia Titan Xp Graphics Card.
The only difference concerning the CNN architecture between the present work and \cite{liu2018interspeech} is that, instead of using the hand movement temporal segmentation derived by the simple procedure (i.e., the left temporal boundary of each phoneme is forced to extend to the left boundary of its previous phoneme), in this work, we use the HPM-based temporal segmentations of hand position and hand shape movement for CNN training.

\paragraph{Hand position feature extraction}
 
In \cite{heracleous2010cued, heracleous2012continuous}, hand position was tracked by detecting the color marks on the hand, and no automatic method was used. In our previous work \cite{liu2018interspeech}, ABMMs were used to automatically extract the hand position in CS on raw hand ROIs.
It is then processed by a simple feed-forward ANN, which contains two standard fully connected layers with ReLU activation function. Two hidden layers with four neurons in each layer are used for the hand position feature processing. After the softmax layer, it will output the posterior probabilities of the target classes with dimension 6, which is the final hand position feature. A mini-batch gradient descent algorithm based on the \emph{Root Mean Square Propagation} (RMSprop)
adaptive learning rate method is used for the parameter optimization in ANN. 

\subsubsection{MSHMM-GMM phonetic decoding}
\label{sec:HMM-GMM phonetic decoding}
 
As in the $\mathcal{S}_3$ architecture \cite{liu2018interspeech}, in the proposed $\mathcal{S}_{\rm re}$ architecture, each phoneme is modeled by a context-dependent triphone MSHMM (i.e., takes into account the contextual information about the preceding and following phonemes) \cite{young1994tree}.
Three emitting states are used with GMM to model the features of lips, hand position and hand shape together with their first derivatives.
The main difference between $\mathcal{S}_3$ and $\mathcal{S}_{\rm re}$ is that
MSHMM-GMMs are used to model the re-synchronized multi-modal features in $\mathcal{S}_{\rm re}$, while in $\mathcal{S}_3$, MSHMM-GMMs are used to model the asynchronous multi-modal features. In $\mathcal{S}_{\rm re}$, the emission probability at state $j$ is
\small
\begin{equation}
b_{j}(\vect{O}^{\rm (LPS)}_{t}) = \prod_{s=1}^{S_{0}}\left [ \sum_{m=1}^{M_{\rm s}}c_{\rm jsm}\textit{N}(\vect{O}^{\rm (LPS)}_{\rm st}; \mu _{\rm jsm};\Sigma_{\rm jsm}) \right ]^{\lambda _{\rm s}},
\end{equation}
\normalsize
where $\vect{O}^{\rm (LPS)}_{\rm t}=[\vect{O}^{\rm (L)^\T},O_{\rm resy}^{\rm (P)^\T},O_{\rm resy}^{\rm (S)^\T}]^\T$ is the merged feature at time $t$, and $S_{0}$ is the number of feature streams, which is set to be $3$ in this work. $\textit{N}(O^{\rm (LPS)}_{\rm st}; \mu _{\rm jsm};\Sigma_{\rm jsm})$ is the probability density function of the observation $O^{\rm (LPS)}_{\rm st}$ for feature stream $s$ and time $t$. $\mu_{\rm jsm}$ is the mean value and $\Sigma_{\rm jsm}$ is the covariance matrix of the Gaussian probability density function with state $j$, mixture Gaussian component $m$ and stream $s$. For stream $s$, $M_{\rm s}$ gaussians are used to model the observation, with weight $c_{\rm jsm}$. In the experiment, $M_{\rm s}$ iteratively increases from 1 to 4 in the training, and is finally set to be $2$ which achieves the best performance. $\lambda_{\rm s}$ is the weight for feature stream $s$, which is optimized using the cross-validation. These weights should satisfy
\begin{equation}
\sum_{s=1}^{3}\lambda _{\rm s} =1, \ \ 0 \leq \lambda _{\rm s}\leq  1.
\end{equation}
Finally, the optimal weights are set to be $0.4$, $0.4$ and $0.2$ for lips, hand shape and hand position, respectively. 

It should be mentioned that, in this work, we do not take into account the pronunciation dictionary or language model in order to directly compare the CS recognition results with the state-of-the-art \cite{liu2018interspeech}, which does not incorporate them. 

\section{Results and Discussions}
\label{sec:Experiment and Discussion}
In this section, we first evaluate the performance of HPM for vowels. Then a two-step evaluation of the proposed CS recognition architecture is carried out to evaluate the performance of the proposed re-synchronization procedure.

\begin{table*}
      \centering
      \small\addtolength{\tabcolsep}{-4pt}
          \caption{Hand position recognition results using multi-Gaussian classifier based on different temporal segmentations of hand position movement for four CS speakers. Ground truth target finger is used as the hand position. Audio-based segmentation is the temporal segmentation used in the previous work \cite{heracleous2010cued, heracleous2012continuous} and the HPM-based segmentation is the temporal segmentation predicted by the proposed HPM for hand movement.}
% %          Confidence interval $\Delta_{95\%}$ is about $4\%$.
              \label{tab:hand_position_4_cases}
          \begin{tabular}{c| c |c |c} 
           \hline\hline
     Recognition accuracy (\%) & audio-based segmentation\ \ \ & HPM-based segmentation \ \ \  & ground truth segmentation \\ [0.4ex] 
    %         \hline
    %  ABMMs  & 45.41 & 54.4 & 62.26 \\    
       \hline
       \textit{LM} & 64.23 & 82.35 & 96.93 \\ 
       \hline
       \textit{SC} & 75.11 & 83.93 & 99.91 \\ 
       \hline
       \textit{MD} & 65.77 & 73.10 & 98.26 \\ 
              \hline
       \textit{CA} & 54.61 & 65.03 & 92.02 \\ 
       \hline\hline
     \end{tabular}
 \end{table*}
\subsection{Evaluation of the HPM-based temporal segmentation of hand position}
\label{subsec:Evaluation of hand preceding model for hand position}
To evaluate the HPM-based temporal segmentation of hand position, we compare it with both the ground truth and audio-based segmentations. (1) For the ground truth, the target instants of all vowels are determined manually, which constitutes a golden reference for the hand position recognition in CS. (2) In the literature  \cite{heracleous2010cued,heracleous2012continuous}, the temporal segmentation of hand position is always based on the audio signal. It is  important to compare this segmentation with the HPM-based segmentation to see the potential benefits of HPM.
 
We first evaluate the proposed HPM by visualizing the hand position distributions in a 2D image using different temporal segmentations. Then, we apply the simple Gaussian classifier to hand position recognition.
 
\begin{figure*}[htbp!]
\minipage{0.33\textwidth}
  \includegraphics[width=\linewidth]{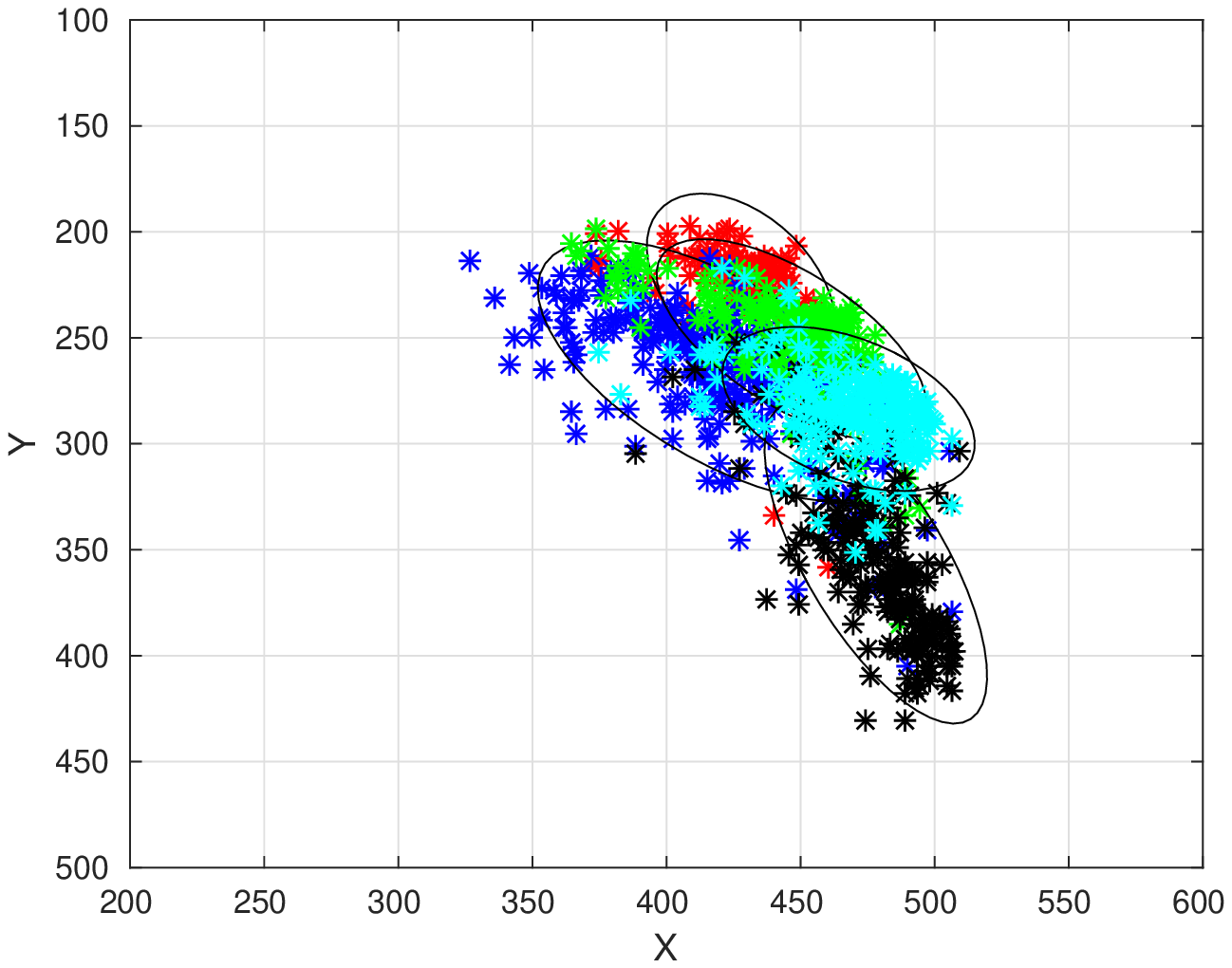}
 \centerline{(a)}\medskip
\endminipage\hfill
\minipage{0.33\textwidth}
  \includegraphics[width=\linewidth]{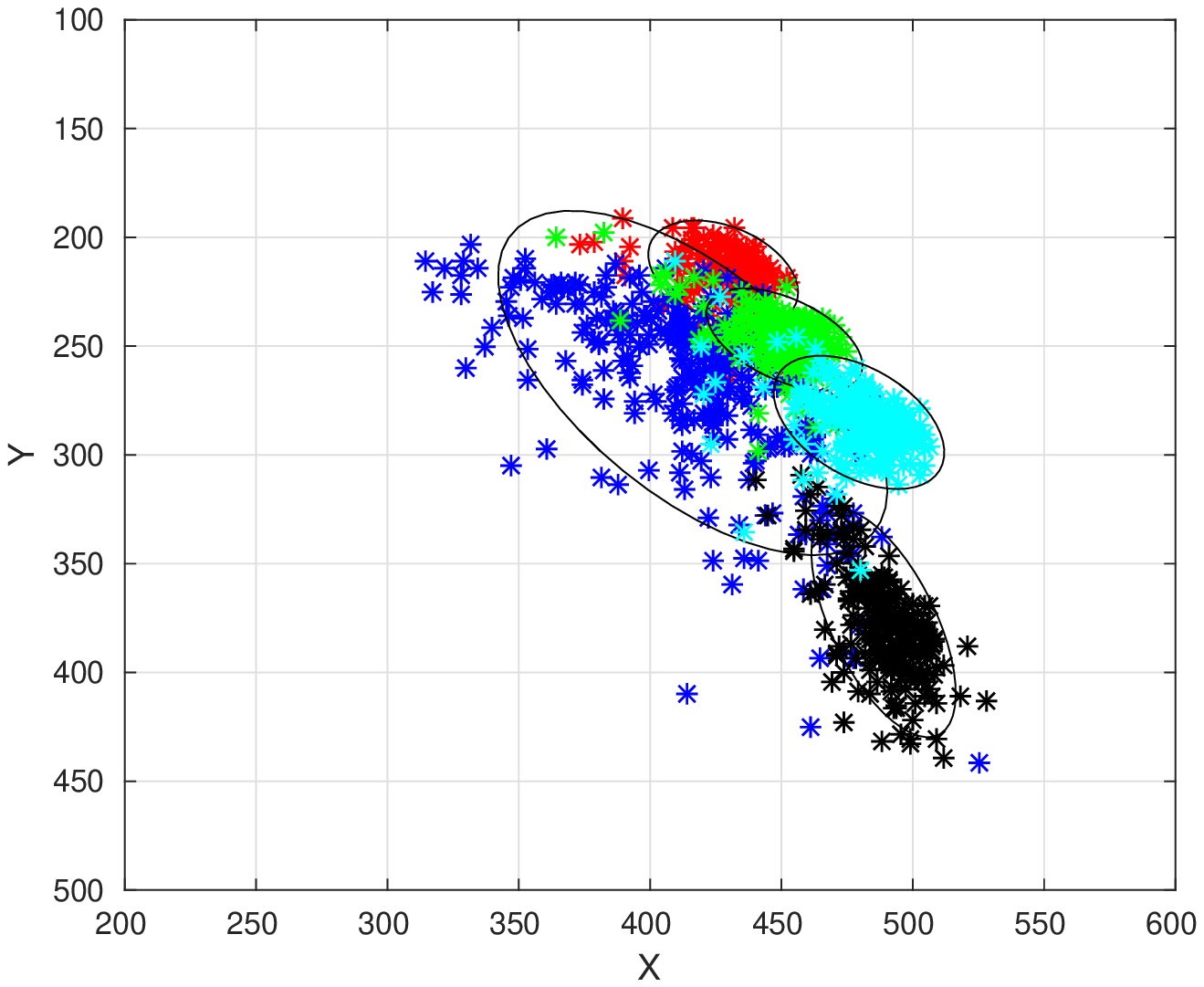}
  \centerline{(b)}\medskip
\endminipage\hfill
\minipage{0.33\textwidth}
  \includegraphics[width=\linewidth]{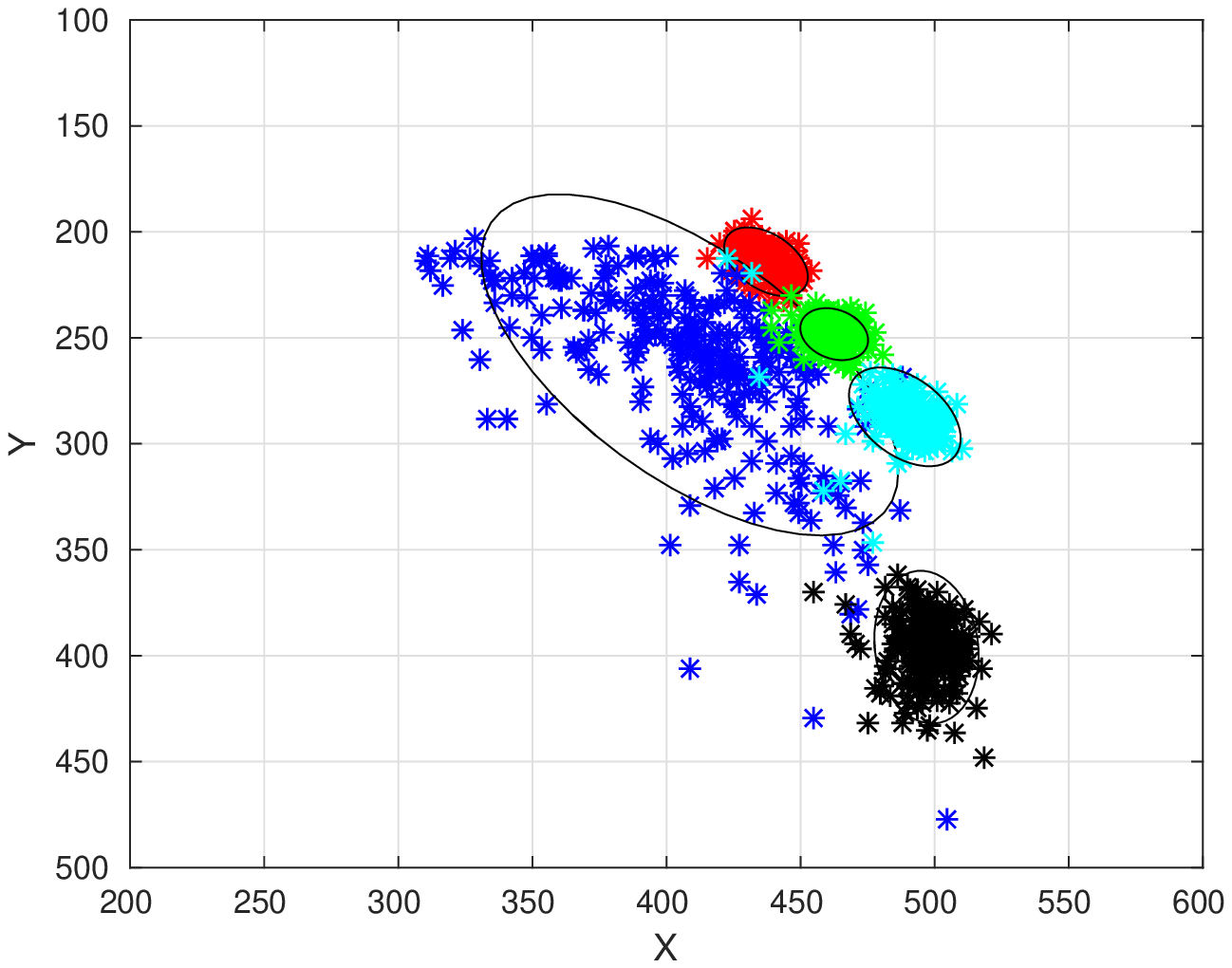}
  \centerline{(c)}\medskip
\endminipage
\\
\minipage{0.33\textwidth}
  \includegraphics[width=\linewidth]{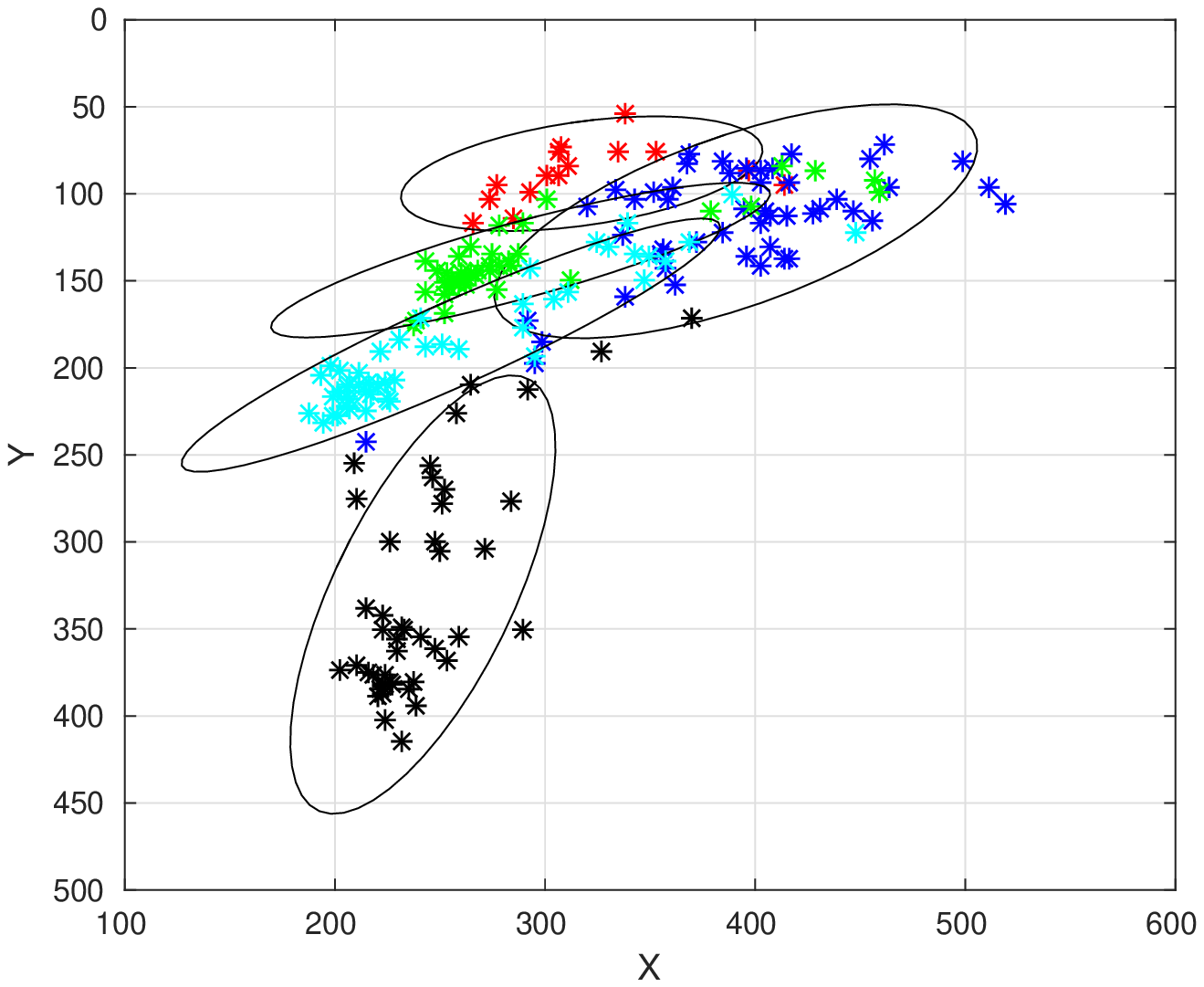}
 \centerline{(d)}\medskip
\endminipage\hfill
\minipage{0.33\textwidth}
  \includegraphics[width=\linewidth]{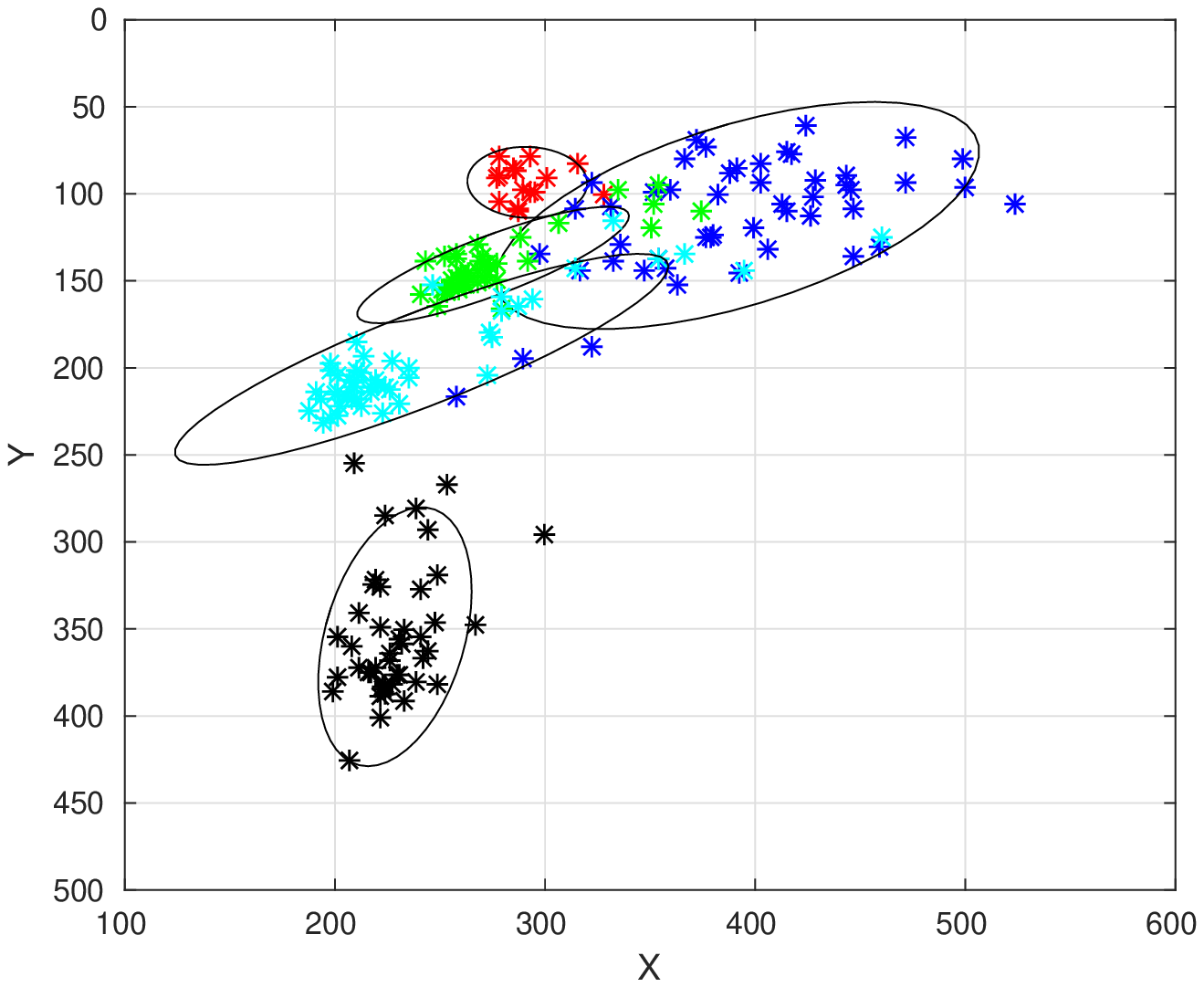}
  \centerline{(e)}\medskip
\endminipage\hfill
\minipage{0.33\textwidth}
  \includegraphics[width=\linewidth]{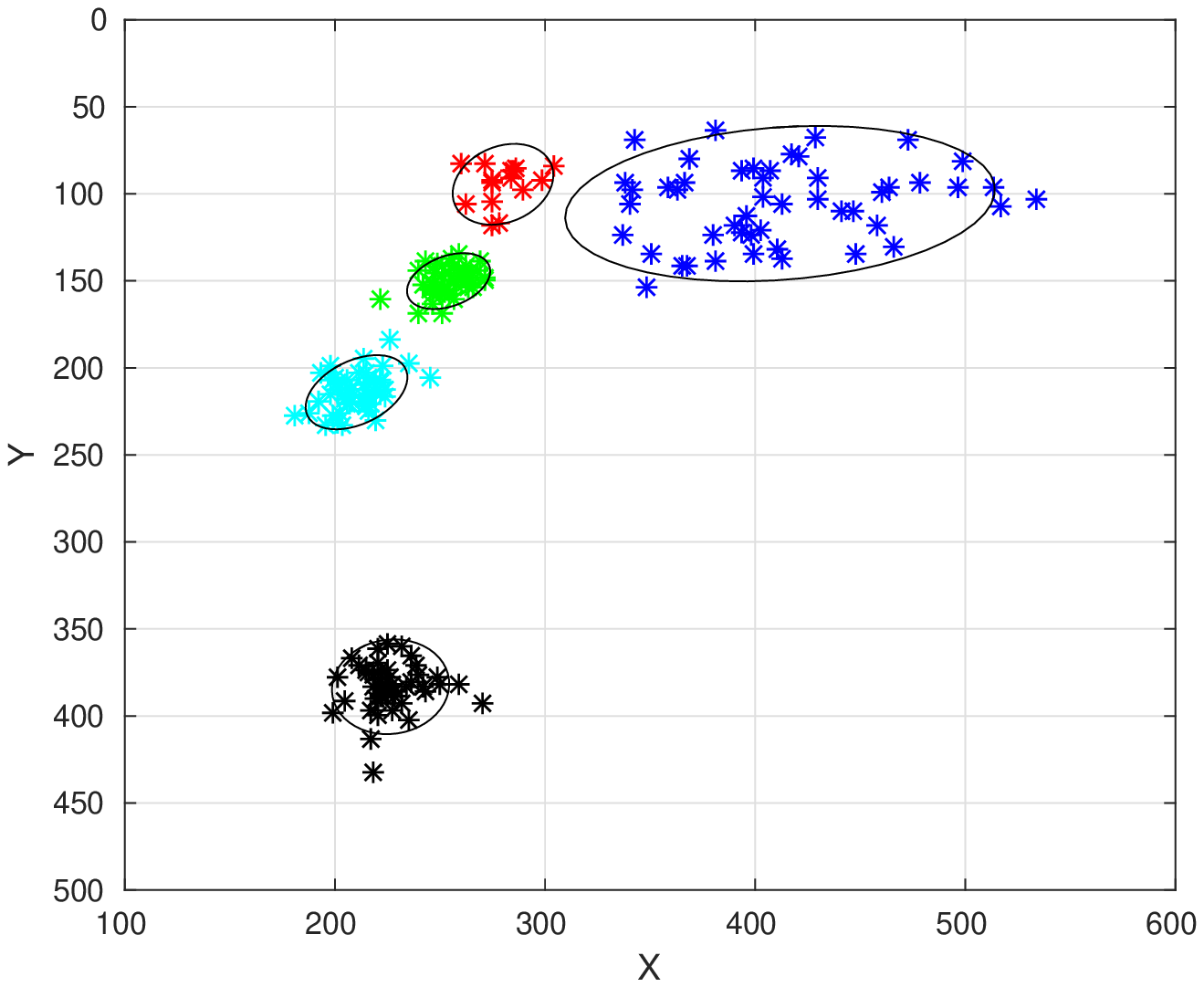}
  \centerline{(f)}\medskip
\endminipage
\\
\minipage{0.33\textwidth}
  \includegraphics[width=\linewidth]{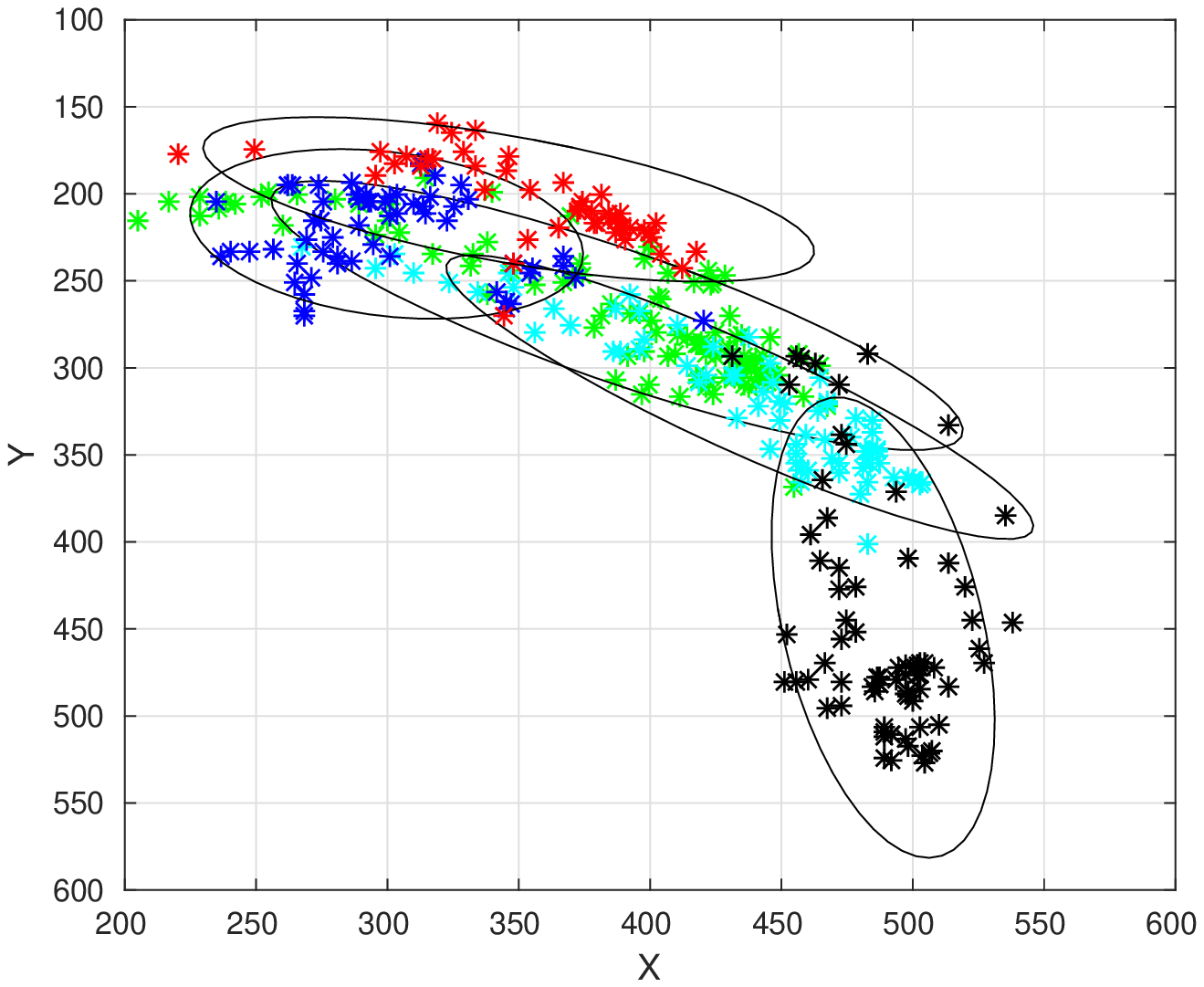}
 \centerline{(g)}\medskip
\endminipage\hfill
\minipage{0.33\textwidth}
  \includegraphics[width=\linewidth]{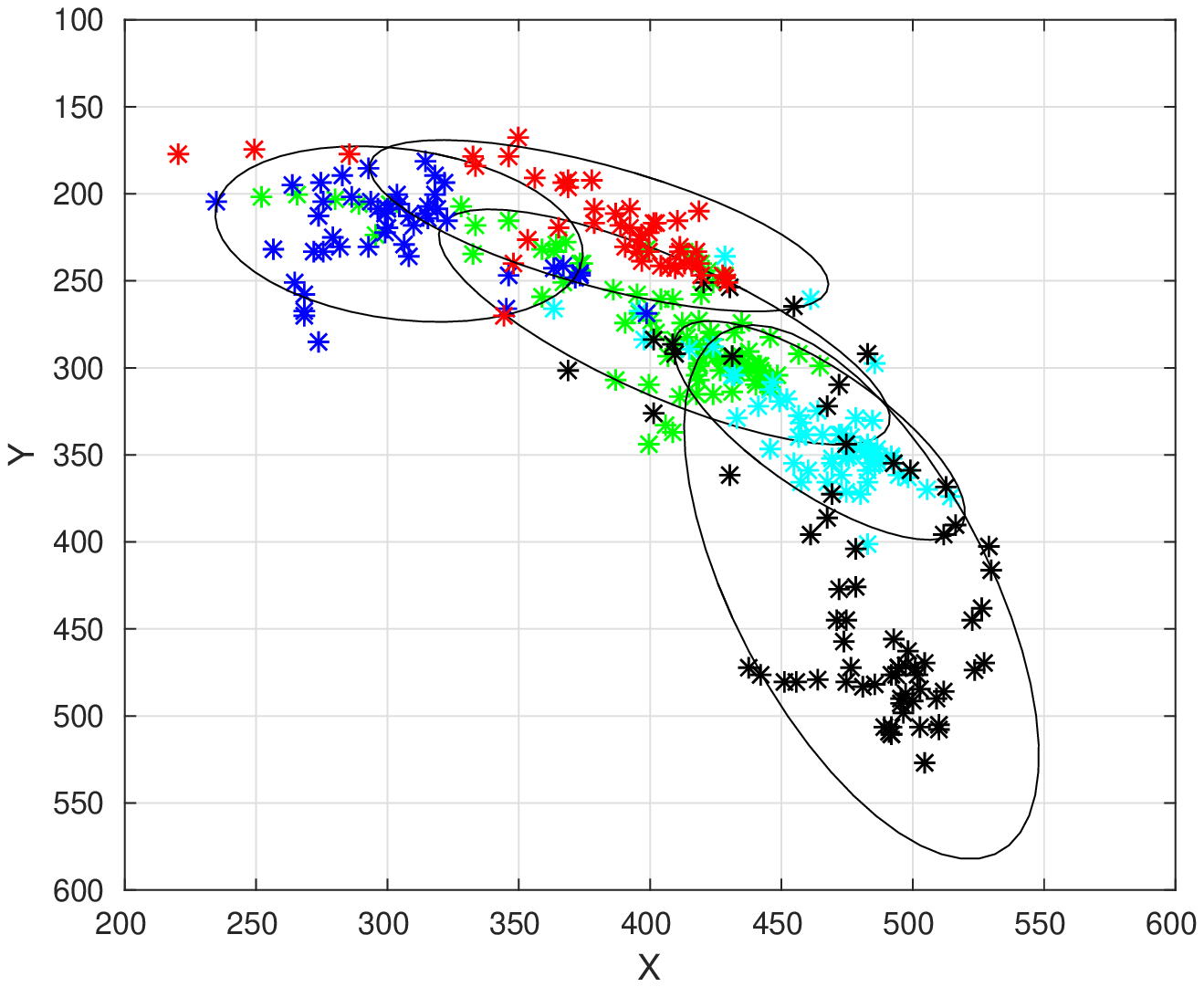}
  \centerline{(h)}\medskip
\endminipage\hfill
\minipage{0.33\textwidth}
  \includegraphics[width=\linewidth]{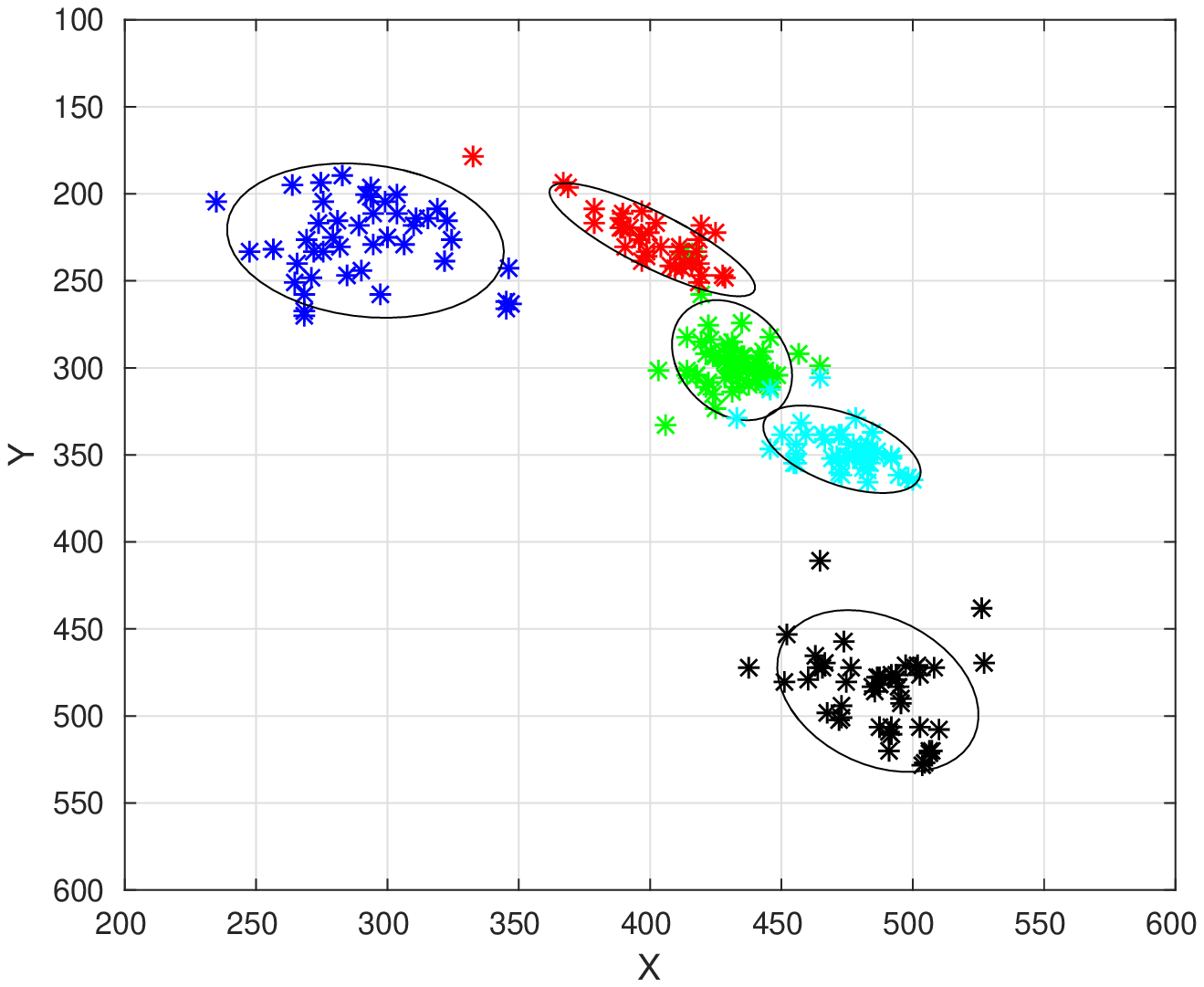}
  \centerline{(i)}\medskip
\endminipage
\\
\minipage{0.33\textwidth}
  \includegraphics[width=\linewidth]{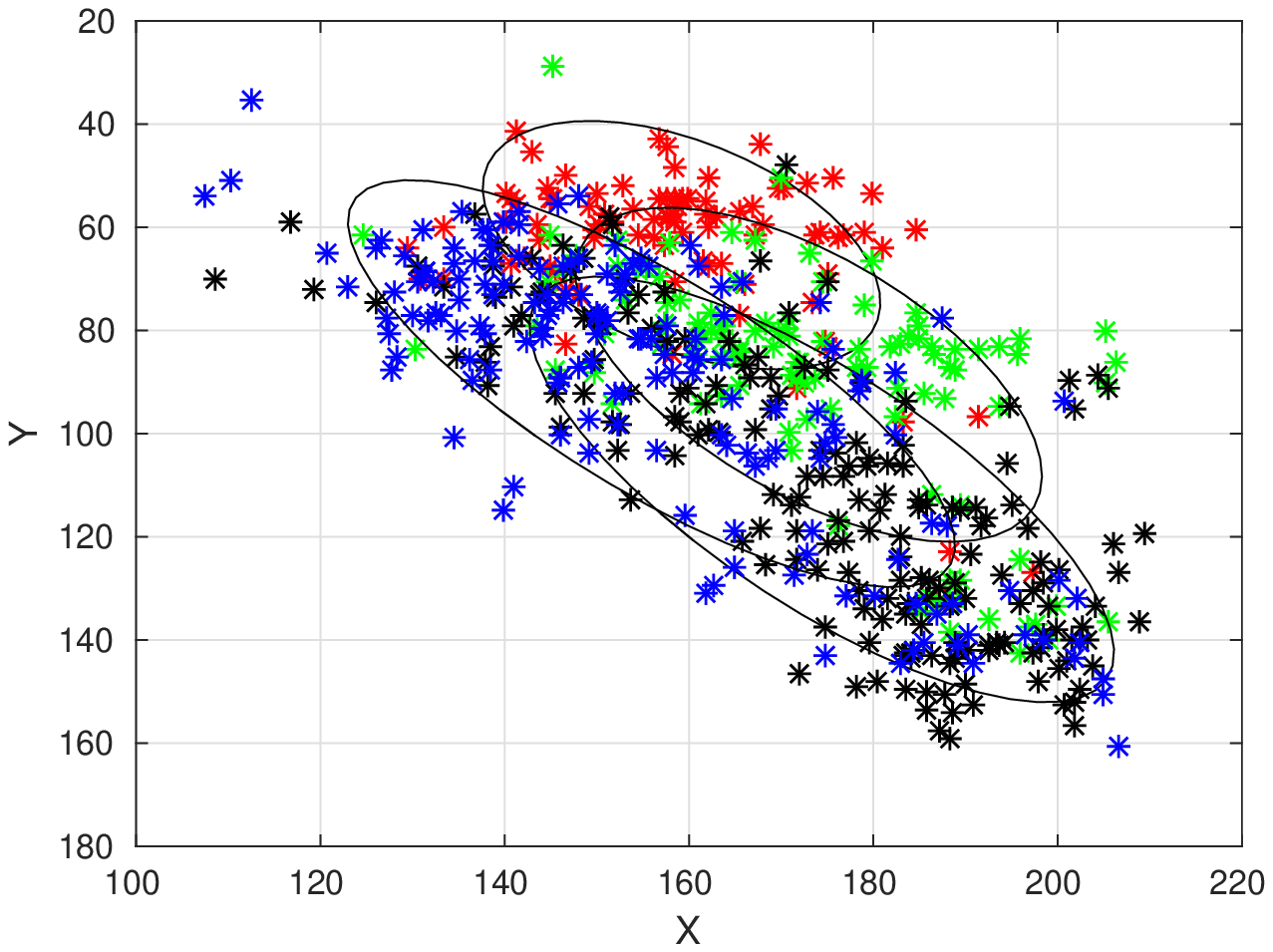}
 \centerline{(j)}\medskip
\endminipage\hfill
\minipage{0.33\textwidth}
  \includegraphics[width=\linewidth]{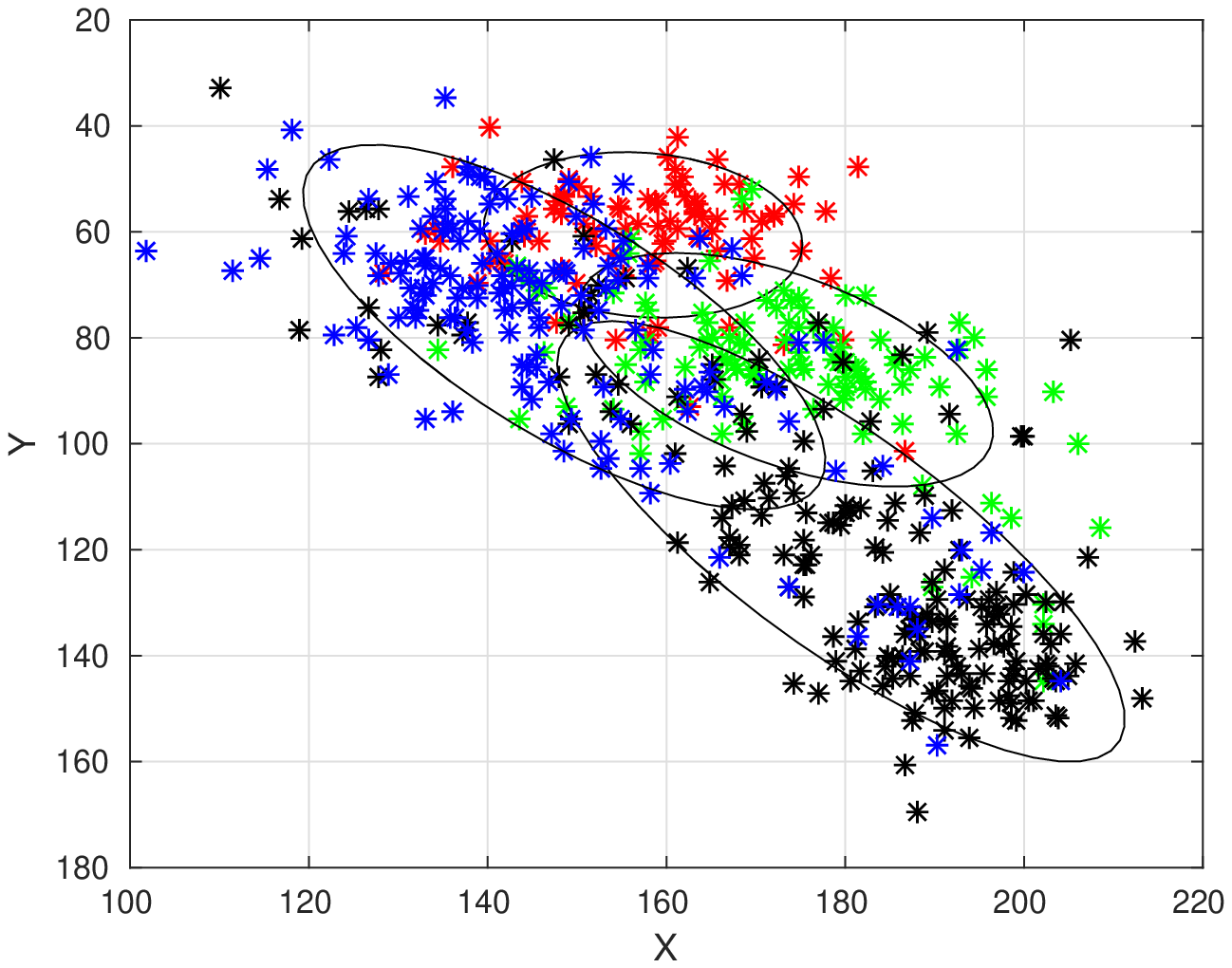}
  \centerline{(k)}\medskip
\endminipage\hfill
\minipage{0.33\textwidth}
  \includegraphics[width=\linewidth]{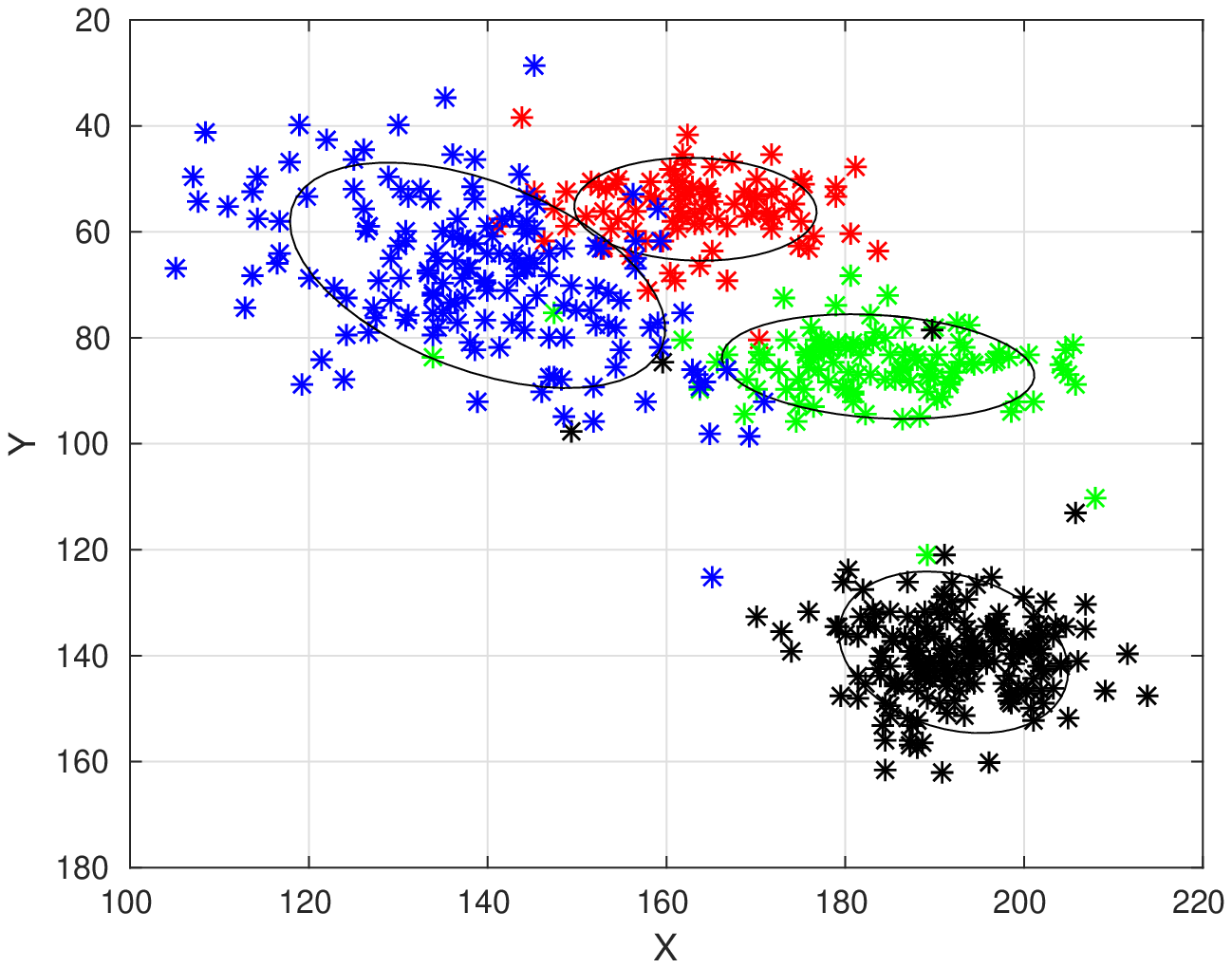}
  \centerline{(l)}\medskip
\endminipage
\caption{Hand position (target finger position) distributions with different temporal segmentations for four CS speakers \textit{LM}, \textit{SC}, \textit{MD} and \textit{CA}, respectively.
(a), (d), (g) and (j): the audio-based segmentation. (b), (e), (h) and (k): the HPM-based segmentation. (c), (f), (i) and (l): the ground truth temporal segmentation. For French CS speakers \textit{LM}, \textit{SC} and \textit{MD}, five groups of points correspond to different hand positions. Red points: \emph{cheek}; green points: \emph{mouth}; black points: \emph{throat}; cyan points: \emph{chin}; blue points: \emph{side}. Note that the speaker \textit{SC} uses her left hand to code CS, while other three use their right hand. In British English CS, only four hand positions are used to code all vowels. Thus, for the speaker \textit{CA}, the meanings of four colors are listed as following. Red points: \emph{mouth}; black points: \emph{throat}; green points: \emph{chin}; blue points: \emph{side}.}
\label{fig:temporal_}
\end{figure*}

Hand position spatial distributions of vowels for four CS speakers using the target finger points and different temporal segmentations are shown in \cref{fig:temporal_}. It is reasonable to assume that the better temporal segmentation corresponds to more distinguishable and separable hand position distributions. 
It can be seen that the Gaussian ellipse becomes more and more distinguishable from left to right for all these four speakers. Note that the second speaker \textit{SC} codes CS using her left hand while other three speakers use their right hand.
Taking the \textit{LM} case corresponding to (a), (b) and (c) as an example,
the points in (a) have large parts of overlaps for five positions, while the points in (b) are significantly separable. In particular, the points at the \emph{throat} and \emph{chin} positions are divided. Moreover, the distribution of the Gaussian ellipse in (b) is very close to that in (c) with only a few intersections between the ellipses.
These distributions efficiently illustrate a satisfactory performance of the HPM.

For further evaluation, we apply these three temporal segmentations to the hand position recognition experiments for four speakers. The experiment results are reported in \cref{tab:hand_position_4_cases}. 
Five multi-gaussian models are trained for the five positions based on a sub-database containing 138 sentences from \textit{LM} corpus, a sub-database containing 44 sentences from \textit{SC} corpus and all vowels from 50 words of \textit{MD} corpus. For the British English CS corpus \textit{CA}, four multi-gaussian models are trained for the four positions based on the \textit{CA} corpus.

Using the target finger positions and the ground truth temporal segmentation, recognition accuracies of all four speakers are higher than other two cases which use the audio-based and HPM-based segmentations. More precisely, for speakers \textit{LM}, \textit{SC}, \textit{MD} and \textit{CA}, the highest accuracies of 96.9\%, 99.9\%, 98.3\% and 92.02\%, are achieved (see \cref{fig:temporal_}(c), (f), (i) and (l)), respectively.
This constitutes golden references to the hand position recognition in CS for these four speakers. 
However, when the audio-based segmentation is used, the recognition accuracies become lower 64.23\%, 75.1\%, 65.8\% and 54.61\% (see \cref{fig:temporal_}(a), (d), (g) and (j)), respectively. When using the proposed HPM-based segmentation, compared with the audio-based segmentation, the accuracies increase significantly to 82.4\%, 83.9\%, 73.1\% and 65.03\% (see \cref{fig:temporal_}(b), (e), (h) and (k)), respectively. All these results show that the audio-based segmentation is indeed not suitable for the hand position recognition, and our proposed HPM improves the performance of hand position recognition significantly. 

Note that for \textit{MD} corpus, the improvement (around 7.5\%) using the HPM-based temporal segmentation is not as evident as \textit{LM} and \textit{SC} corpora. This is because the \textit{MD} corpus is composed of single words with a comparably short time duration, while \textit{LM} and \textit{SC} corpora are made of continuous sentences. Therefore the hand preceding phenomenon in \textit{MD} corpus is not as strong as other two corpora.

Above all, we see that the HPM-based temporal segmentation can significantly improve the performance of hand position recognition.

\subsection{Evaluation of the re-synchronization procedure applied to the vowel and consonant recognition in CS}
\label{subsec: Evaluation of the re-synchronization procedure applied to the vowel and consonant recognition in Cued Speech}

Now, we evaluate the proposed re-synchronization procedure by carrying out the first step, i.e., vowel and consonant recognition experiments based on the fused features of two streams.
We use MSHMM-GMM to see the benefits of this re-synchronization procedure. 

\subsubsection{Vowel recognition based on the fusion of lips and hand position}
\label{sec:Vowel Cued Speech recognition}

We mainly discuss the vowel recognition based on the fusion of lips and hand position. 
The vowel recognitions using only lips and only hand position information are also presented. 

Lips and hand position features are merged and fed to a two-stream MSHMM with triphone context-dependent modeling. 14 vowels (i.e., [i, e, \textipa{E}, a, y, \textipa{\o}, u, o, \textipa{O}, \textipa{\~E}, \textipa{\~\oe}, $\tilde{{\rm a}}$, \textipa{\~O}, \textipa{@}]) plus \emph{silence} are the target labels. Results are shown in \cref{tab:Vowel recognition using lips and hand position in CS}. We see that the vowel recognition using the re-synchronization procedure \eqref{equ:vowel_140} obtains a higher correctness (74.65\%) than the one without using it (70.12\%). 
An improvement of about 4.5\% is achieved.
For the single stream case, using only lips and only hand position give very similar and low correctness. 
This is reasonable since these two streams are supposed to be combined to recognize the vowels, and every single stream only carries the viseme-level information.

 \begin{table}[h!]
     \centering
     \small\addtolength{\tabcolsep}{-8pt}
         \caption{Vowel recognition results using lips and hand position. ${\rm T}_{\rm corr}$ is the correctness and `---' means that this case does not exist. The shorthand `one-stream' means the CS vowel recognition using only one stream feature. `non-resyn' means the CS recognition without using the re-synchronization procedure, while `resyn' means using this procedure.}
         \label{tab:Vowel recognition using lips and hand position in CS}
         \scalebox{1.1}{\setlength{\tabcolsep}{1mm}{
          \begin{tabular}{l c c c}
         \\  \hline\hline
       \rule{0pt}{6pt} ${\rm T}_{\rm corr}$ & one-stream  & non-resyn & resyn\\
      \hline
      \rule{0pt}{6pt} only lips & 34.88\% & ---& --- \\
      \hline
   \rule{0pt}{6pt}  only hand position & 35.71\% & --- & --- \\
      \hline
  \rule{0pt}{6pt} lips + hand position &---& 70.12\% & 74.65\% \\% [0.3ex]
    \hline\hline
    \end{tabular}}}
\end{table}

\subsubsection{Consonant recognition based on the fusion of lips and hand shape}
\label{sec:Consonant Cued Speech recognition}
 Lips and hand shape features are merged and fed to a two-stream MSHMM-GMM with triphone context-dependent modeling. 18 consonants (i.e., [p, t, k, b, d, f, s, \textipa{S}, v, z, \textipa{Z}, m, n, l, g, r, j, w]) plus \emph{silence} are the target labels.
In \cref{tab:Consonant recognition using lips and hand shape in CS}, we can see that using the re-synchronization procedure achieves a higher correctness (82.28\%) than the one without using it (80.33\%). For the single stream case, using only lips and only hand shape obtain a much lower correctness compared with the fused feature. 

 \begin{table}[h!]
     \centering
    \small\addtolength{\tabcolsep}{-8pt}
         \caption{Consonant recognition using lips and hand shape. ${\rm T}_{\rm corr}$ is the correctness and `---' means that this case does not exist. The shorthand `one-stream' means the CS consonant recognition using only one stream feature. `non-resyn' means the CS recognition without using the re-synchronization procedure, while `resyn' means using this procedure.}
         \label{tab:Consonant recognition using lips and hand shape in CS}
       \scalebox{1.1}{\setlength{\tabcolsep}{1mm}{
          \begin{tabular}{l c c c}
          \hline\hline
       \rule{0pt}{4pt} ${\rm T}_{\rm corr}$ & one-stream & non-resyn & resyn\\
      \hline
        \rule{0pt}{4pt}   only lips & 42.25\% & --- & --- \\
      \hline
       \rule{0pt}{4pt}    only hand shape & 56.99\% & --- & --- \\
      \hline
      \rule{0pt}{4pt} lips +  hand shape & --- & 80.33\% & 82.28\% \\% [0.3ex]
    \hline\hline
    \end{tabular}}}
\end{table}

\subsection{Evaluation of the re-synchronization procedure applied to the automatic CS phoneme recognition}
\label{subsec:Results of the automatic recognition using the re-synchronization procedure}
\begin{figure*}[htbp!]%[h!]
\begin{minipage}[b]{1.0\linewidth}
\centering
\centerline{\includegraphics[width=12.3cm]{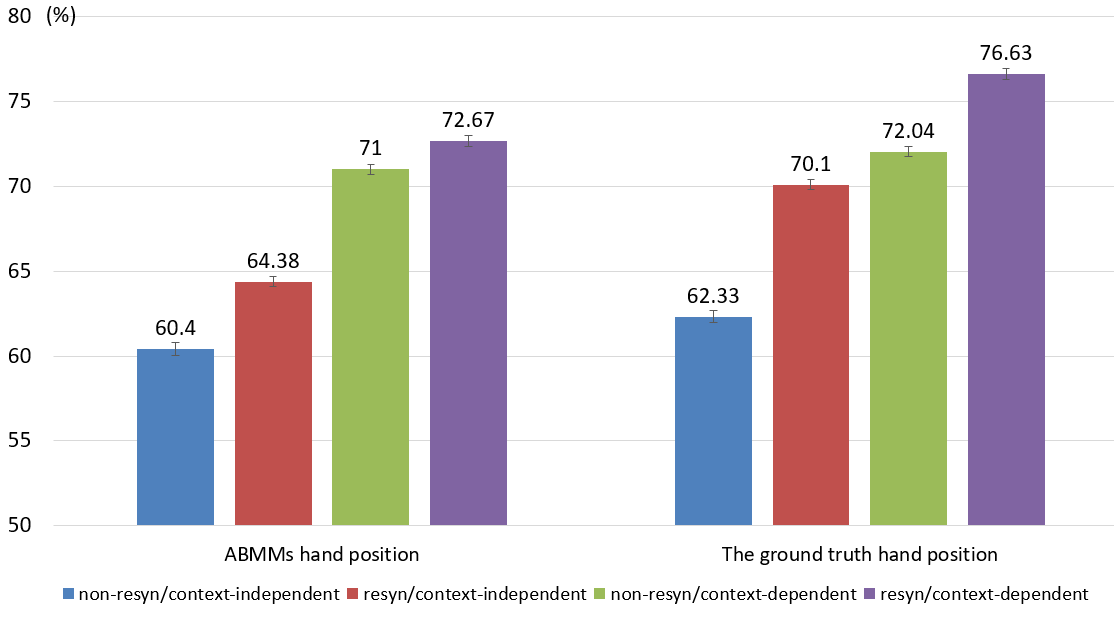}}
\end{minipage}
\caption{Ablation study of the continuous CS phoneme recognition with and without using the proposed re-synchronization procedure and the context-dependent modeling. The shorthand `non-resyn' means the CS recognition without using the re-synchronization procedure, while resyn means using the re-synchronization procedure. The error bars are the stds of the results (less than 0.5) and all the differences are statistically significant.}
\label{fig:aligned_feature_comparsion}
\end{figure*}

To further evaluate the proposed re-synchronization procedure, a new multi-modal CS phoneme recognition architecture $\mathcal{S}_{\rm re}$ is investigated (see \cref{fig:HMM_architecture_resychrony}). We compare the result of $\mathcal{S}_{\rm re}$ with the state-of-the-art $\mathcal{S}_3$ in \cite{liu2018interspeech}, which does not take into account the asynchrony issue in CS. The results are shown in \cref{fig:aligned_feature_comparsion}.

Now we make some analysis.
We first focus on the case when the hand positions given by ABMMs are used, as this is the case in \cite{liu2018interspeech}. 
Using $\mathcal{S}_3$ architecture, the phoneme recognition correctness is 71.0\%, without using any re-synchronization procedure. When the proposed re-synchronization procedure is incorporated, the recognition correctness increases to 72.67\%, which shows a minor improvement of 1.67\% (see columns 3 and 4 in \cref{fig:aligned_feature_comparsion}). 
We realize that in $\mathcal{S}_{\rm re}$, the triphone context-dependent modeling of MSHMM is helpful to correcting the recognition errors, which could be the co-articulation or the asynchrony of multi-modalities \cite{schwartz2009multimodal}.
Therefore, using the context-dependent modeling may hide the effect of the re-synchronization procedure. In order to avoid this phenomenon, we examine the phoneme recognition correctness without using this context-dependent modeling. Then only 60.4\% is obtained without any re-synchronization procedure, while it increases to 64.38\% when using the proposed re-synchronization procedure (see columns 1 and 2 in \cref{fig:aligned_feature_comparsion}). It can be seen that the improvement of about 4\% is more evident than the case of using the context-dependent modeling (1.67\%). 

The above improvement is weak. This may be because (1) only a weight of 0.2 is applied to the hand position stream, and (2) the hand position extracted by ABMMs has some errors (see \cref{subsec:Automatic continuous Cued Speech recognition}). 
For the first reason, as mentioned in \cref{sec: Problem formulation_}, the hand position feature is much more sensitive to the asychrony problem in CS. The small weight reduces the importance of the re-synchronization procedure.
For the second reason, the errors directly reduce the efficiency of the re-synchronization procedure and cause some interference of the CS recognition system, since the hand position target can be recognized accurately only if the correct hand position feature is used at a good temporal boundary for a particular vowel. 
Note that in this work, we only consider the interference caused by the hand position errors because the CNN-based features for lips and hand shape are shown to be correct in \cite{liu2018interspeech}. 

In order to see the effects when the hand position is correct, we use the ground truth hand position instead of that given by the ABMMs. 
The CS phoneme recognition results in this case are shown in \cref{fig:aligned_feature_comparsion}, from column 5 to column 8. 

Now we make some discussions.
In this case, without the context-dependent modeling or re-synchronization procedure, the correctness is 62.33\%, which is close to the result of 60.4\% based on the hand positions by ABMMs. This can be explained by the above first reason. 
However, when the re-synchronization procedure is used, a correctness of 70.1\% is achieved, which shows a significant improvement of about 7.7\% (see columns 5 and 6 in \cref{fig:aligned_feature_comparsion}). This shows benefits of the proposed re-synchronization procedure.

When using the context-dependent modeling (see columns 7 and 8 in \cref{fig:aligned_feature_comparsion}), the correctness is 72.04\% without the re-synchronization procedure. 
However, when using both of them, a significant correctness of 76.63\% is obtained (with an improvement of 4.6\%) compared with the state-of-the-art \cite{liu2018interspeech} in the automatic continuous CS phoneme recognition case. 
Moreover, this result also outperforms the state-of-the-art 74.4\% \cite{heracleous2010cued} which is for the automatic isolated CS phoneme recognition. 

We also calculate the CS phoneme recognition correctness when using only the lips information.
The result is low (around 30\%). 
It can be seen that by using the CS, the phoneme recognition performance increases by a large margin of around 46.63\% compared with the case of using lip reading only. 
This confirms that CS can significantly help the Deaf/Hard of hearing on speech perception and production.

We note that the proposed HPM is speaker-dependent, i.e., each speaker has one HPM. We are now trying to employ more Deaf/Hard of hearing CS cuers and collect more data to generalize this method to be speaker-independent. Indeed, the proposed automatic CS recognition framework in this work is speaker-independent.

\section{Conclusion}
In this work, a novel re-synchronization procedure is proposed for the CS multi-modal feature fusion applied to a practical automatic continuous French CS recognition system (i.e., CNN-MSHMMs). First, by exploring the relationship between the HPT and the time instants of phonemes in French continuous sentences, we obtain the optimal HPTs for all vowels ($140$ms) and for all consonants ($60$ms), and develop new HPMs.
Then, the re-synchronization procedure is proposed by delaying the hand position and shape streams with these two different optimal HPT, respectively. The lips and hand features can be re-synchronized on average. By incorporating the proposed re-synchronization procedure to the CNN-MSHMMs based CS recognition system, we propose a new architecture $\mathcal{S}_{\rm re}$ that takes into account the asynchrony issue in the CS recognition. The evaluation of the automatic CS phoneme recognition using $\mathcal{S}_{\rm re}$ achieves a significant improvement (about 4.6\%) compared with the state-of-the-art architecture $\mathcal{S}_3$.
In the future study, the fusion method for the asynchronous multi-modalities is a very interesting and challenging research direction. Besides, a speaker-independent multilingual (e.g., French, English and Mandarin Chinese CS \cite{liu2019MCCSpilot}) CS fusion method is also worthy to be investigated.

\section*{Acknowledgement}
The authors would like to thank the CS speakers for their time spent on the
French CS data recording, 
and Thomas Hueber for his help in CNN-HMM. 
The authors would like to thank the referees for their valuable comments and suggestions. This work is supported by a PhD thesis grant of Université Grenoble Alpes in France, and in part by the Natural Sciences and Engineering Research Council of Canada under Grant RGPIN239031.

% \ifCLASSOPTIONcaptionsoff
%   \newpage
% \fi

\bibliographystyle{liubib1}
\bibliography{referenceli}

\begin{IEEEbiography}[{\includegraphics[width=1in,height=1.25in,clip,keepaspectratio]{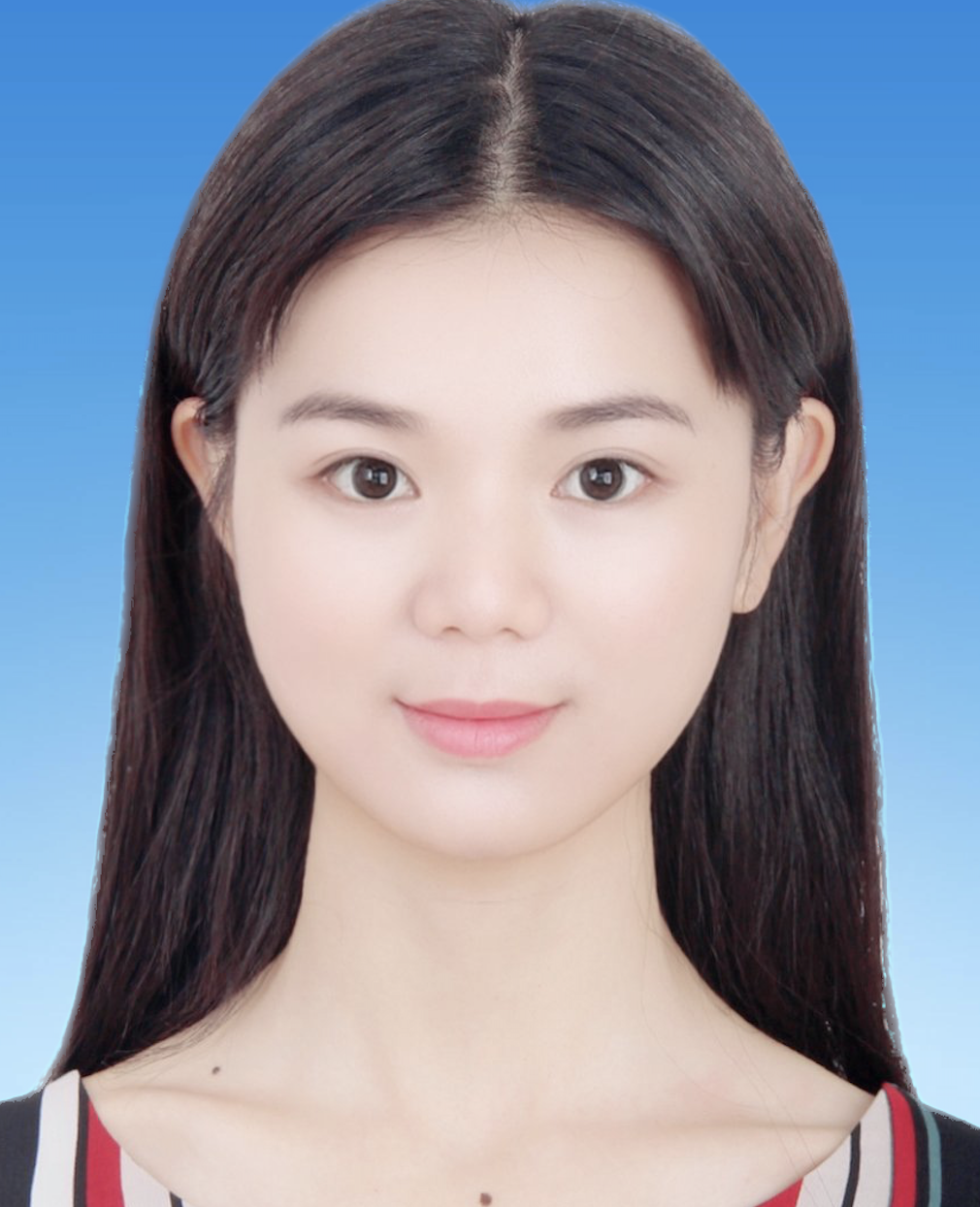}}]{Li Liu}
received the B.S. degree in 2014 from Mathematics and Applied Mathematics department of Heilongjiang University, Harbin, China, the M.S. degree in 2015 from Master of Science in Industrial and Applied Mathematics program in the Université Grenoble Alpes, Grenoble, France, and the Ph.D. degree in 2018 from Gipsa-lab, Université Grenoble Alpes, Grenoble, France. From September 2018 to September 2019, she was a postdoc researcher in Department of Electrical, Computer and Biomedical Engineering, Ryerson University, Toronto, Canada. Now, she works as a research scientist in Shenzhen Research Institute of Big Data, Shenzhen, China.

Her current research interests include automatic audio-visual speech recognition, multi-modal fusion, Cued Speech development, lips/hand gesture recognition and medical imaging.
\end{IEEEbiography}

\begin{IEEEbiography}[{\includegraphics[width=1in,height=1.25in,clip,keepaspectratio]{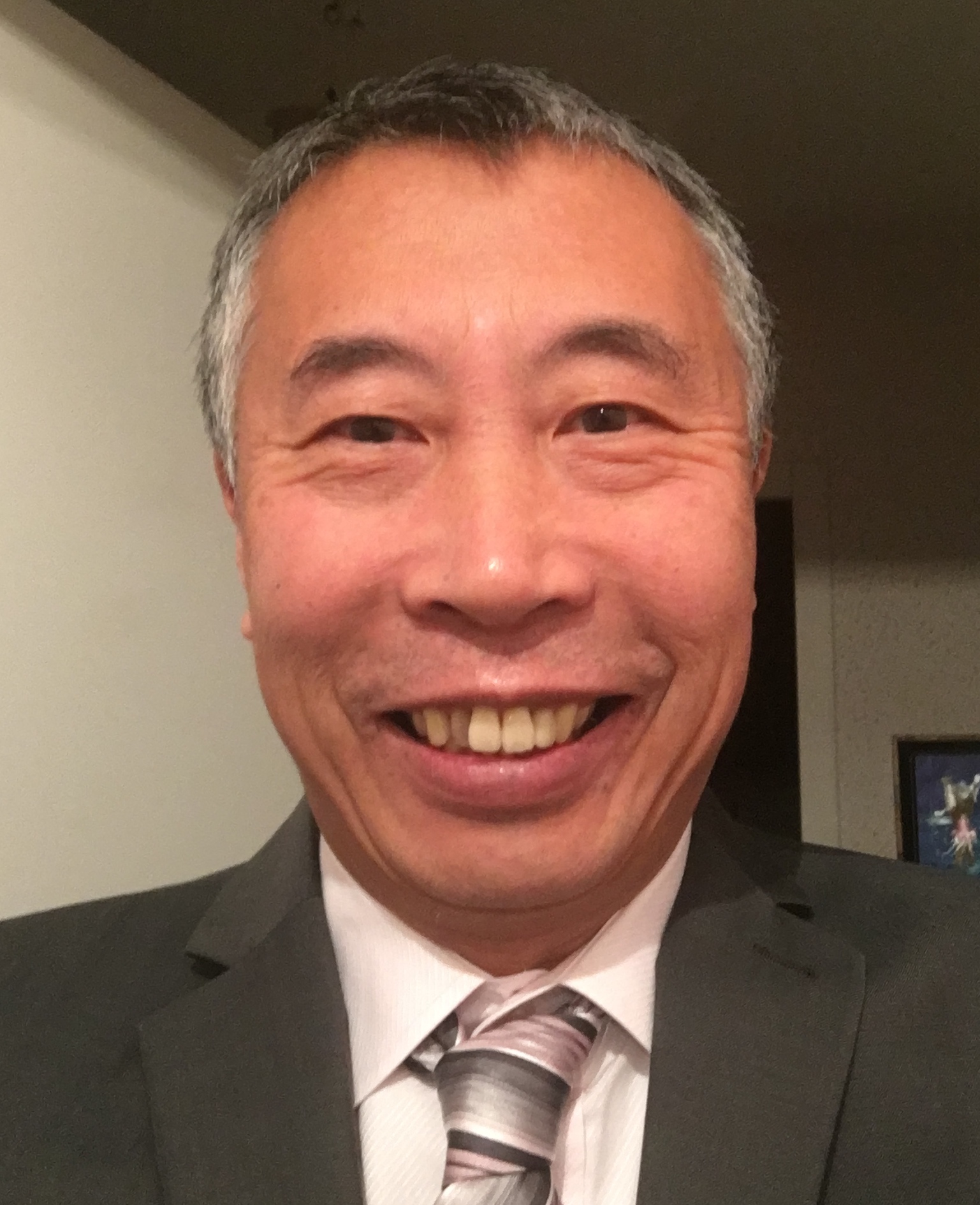}}]{Gang Feng}
received the B.S. degree from Huazhong Institute of Technology, Wuhan, China, in 1982, and the M.S. and Ph.D. degrees from Grenoble Institute of Technology, Grenoble, France, in 1983 and 1986, respectively.

From 1987 to 1989, he worked as engineer at OROS Campany, Grenoble, France. From 1989 to 1998, he was an associate professor with Ecole Nationale Supérieure d’Electronique et de Radioélectricité de Grenoble, France, where he is a Professor since 1998. He is a professor of GIPSA laboratory (Grenoble Automatics, Signal and Image Processing, Speech Processing). His current research interests include articulatory-acoustic modeling, speech coding, speech synthesis and multi-modal speech analysis.
\end{IEEEbiography}
\newpage
\begin{IEEEbiographynophoto}{Denis Beautemps}
is a speech scientist. He joined the French CNRS National Research Center in 1998 as a researcher. Currently he works at the Grenoble GIPSA-lab, France. Much of his research
is dedicated to multi-modality processing including the gesture component of Cued Speech, which is a
hand coding complement to lip reading for deaf communication. Twenty years ago, he promoted
the pioneer works on Cued Speech under production aspect. Since then, he has managed six
PhD students and three post-doc in the domain of CS and lip reading analysis, synthesis and
recognition. Dr. Denis Beautemps has published in more than eighty international peer-reviewed
journals and conferences.
 \end{IEEEbiographynophoto}
\begin{IEEEbiography}[{\includegraphics[width=1in,height=1.25in,clip,keepaspectratio]{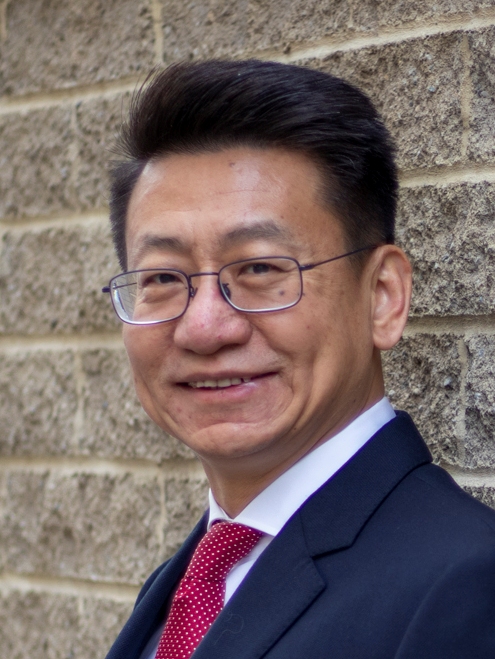}}]{Xiao-Ping Zhang} received B.S. and Ph.D. degrees from Tsinghua University, in 1992 and 1996, respectively, both in Electronic Engineering. He holds an MBA in Finance, Economics and Entrepreneurship with Honors from the University of Chicago Booth School of Business, Chicago, IL. 

Since Fall 2000, he has been with the Department of Electrical, Computer and Biomedical Engineering, Ryerson University, Toronto, ON, Canada, where he is currently a Professor and the Director of the Communication and Signal Processing Applications Laboratory. He has served as the Program Director of Graduate Studies. He is cross-appointed to the Finance Department at the Ted Rogers School of Management, Ryerson University. He was a Visiting Scientist with the Research Laboratory of Electronics, Massachusetts Institute of Technology, Cambridge, MA, USA, in 2015 and 2017. He is a frequent consultant for biotech companies and investment firms. He is the Co-Founder and CEO for EidoSearch, an Ontario-based company offering a content-based search and analysis engine for financial big data. His research interests include image and multimedia content analysis, machine learning, statistical signal processing, sensor networks and electronic systems, and applications in big data, finance, and marketing. 

Dr. Zhang is a registered Professional Engineer in Ontario, Canada, and a member of Beta Gamma Sigma Honor Society. He is the general Co-Chair for the IEEE International Conference on Acoustics, Speech, and Signal Processing, 2021. He is the general co-chair for 2017 GlobalSIP Symposium on Signal and Information Processing for Finance and Business, and the general co-chair for 2019 GlobalSIP Symposium on Signal, Information Processing and AI for Finance and Business. He is an elected Member of the ICME steering committee. He is the General Chair for the IEEE International Workshop on Multimedia Signal Processing, 2015. He is the Publicity Chair for the International Conference on Multimedia and Expo 2006, and the Program Chair for International Conference on Intelligent Computing in 2005 and 2010. He served as a Guest Editor for Multimedia Tools and Applications and the International Journal of Semantic Computing. He was a tutorial speaker at the 2011 ACM International Conference on Multimedia, the 2013 IEEE International Symposium on Circuits and Systems, the 2013 IEEE International Conference on Image Processing, the 2014 IEEE International Conference on Acoustics, Speech, and Signal Processing, the 2017 International Joint Conference on Neural Networks and the 2019 IEEE International Symposium on Circuits and Systems. He is a Senior Area Editor for the IEEE TRANSACTIONS ON SIGNAL PROCESSING and the IEEE TRANSACTIONS ON IMAGE PROCESSING. He was an Associate Editor for the IEEE TRANSACTIONS ON IMAGE PROCESSING, the IEEE TRANSACTIONS ON MULTIMEDIA, the IEEE TRANSACTIONS ON CIRCUITS AND SYSTEMS FOR VIDEO TECHNOLOGY, the IEEE TRANSACTIONS ON SIGNAL PROCESSING, and the IEEE SIGNAL PROCESSING LETTERS. He received 2020 Sarwan Sahota Ryerson Distinguished Scholar Award – the Ryerson University highest honor for scholarly, research and creative achievements. He is awarded as IEEE Distinguished Lecturer for the term from January 2020 to December 2021 by IEEE Signal Processing Society.
\end{IEEEbiography}
\end{document}